\documentclass[A4,twoside,12pt,reqno]{article}

\usepackage[super,compress]{cite}
\usepackage{graphicx}
\usepackage{float}
\usepackage[latin1]{inputenc}

\usepackage{amssymb}
\usepackage{amsthm}
\usepackage[tbtags]{amsmath}
\usepackage{doc}
\usepackage{latexsym}
\usepackage{amscd}
\usepackage[frame,cmtip,arrow,matrix,line,graph,curve]{xy}
\usepackage{graphics}
\usepackage{epsfig}
\usepackage{eucal}
\usepackage{mathrsfs}
\usepackage[margin=1.3in]{geometry}

\begin{document}
\thispagestyle{plain}
 \markboth{}{}
\small{\addtocounter{page}{0} \pagestyle{plain}

\noindent\centerline{\Large \bf Could the Lyra manifold be the hidden source} \\ 
\noindent\centerline{\Large \bf  of the dark energy? }\\

\noindent\centerline{$^1$Kangujam Priyokumar Singh}\\
\noindent\centerline{$^2$Mahbubur Rahman Mollah}\\
\textbf{}
\noindent\centerline{$^1$Department of Mathematical Sciences, Bodoland University,}\\
\noindent\centerline{Kokrajhar, Assam-783370, India. }\\
\noindent\centerline{pk\_mathematics@yahoo.co.in}
\noindent\centerline{$^2$Department of Mathematics, Commerce College,}\\
\noindent\centerline{Kokrajhar, Assam - 783370, India }\\
\noindent\centerline{mr.mollah123@gmail.com}\\

\footnote{Preprint of an article published in [Int. J. Geom. Methods Mod. Phys., Vol. 14, No. 04, 1750063 (2017) DOI: 10.1142/S0219887817500633]\textsuperscript{\textcopyright}[World Scientific Publishing Company] [Journal URL: www.worldscientific.com/worldscinet/ijgmmp]}

\textbf{Abstract:} In the course of investigation of our present universe by considering the five-dimensional locally rotationally symmetric (LRS) Bianchi type-I universe with time-dependent deceleration parameters in  Lyra manifold, it is excitingly found that the geometry itself of Lyra manifold behaves and consistent with present observational findings for accelerating universe. The behavior of the universes and their contribution to the process of evolution are examined. While studying their physical, dynamical and kinematical properties for different cases, it is found that this model is a new and viable form of model universe containing dark energy. It will be very helpful in explaining the present accelerated expansion behavior of the universe.

\section{Introduction}	
Many researchers and Scientists are putting huge effort to explain the dynamics of the universe and to understand the future evolution of the universe from the ancient times. However, till today we are not in a state to provide an exactly clear statement about the origin and evolution of our universe. From different literatures and philosophical point of views we found that different mines provide different opinions about our universe. The universe is full of mysterious elements and numerous effects of interactions which cannot be detected and difficult to explain even with advanced technology. Due to these reasons the most challenging problems in Astrophysics and modem cosmology is to understand the late time acceleration of the universe. In recent years, most of researchers have drawn considerable attention in the context of dark energy and modified theories of gravity to study the various aspects of the universe. On the other, hand many prominent results of the cosmological observations like Type SNeIa supernovae [1-3], CMB (Cosmic Microwave Background) anisotropies [4-5], the large scale galaxies structures of universe [6], Baryon Acoustic Oscillations [7], WMAP (Wilkinson Microwave Anisotropy Probe) [8] Sachs-Wolfe effects [9] and SDSS [10] are noticed to us for the cosmic acceleration with direct and indirect evidence. Not only the above mentioned observations and surveys, but also some other new cosmological results and data sets like Planck [11], Atacama Cosmology Telescope (ACT) [12], South Pole Telescope Sunyaev-Zel'dovice (SPT-SZ) survey [13] have measured the temperature and polarization of  CMB to exquisite precision, are also supporting this fact. These results are more acceptable to the community and beneficial to the researchers to understand about the universe in this modern era.\\
\\
From the recent literatures and findings, we know that dark energy dominates the universe with positive energy density and negative pressure, responsible to produce sufficient acceleration in late time evolution of the Universe. Some of the important claimants of dark energy are tachyons [14], chaplygin gas [15], phantom [16], k-essence and quintessence [17] along with other four elements, i.e. dark matter, baryons, radiation and neutrinos. But so far there is no direct detection of such exotic fluids. Although the literature is now flooded with hundreds of model for dark energy what we lack is precise cosmological data coming from variety of observations involving both background and inhomogeneous universe that can discriminate among these models. In this connection if we accept that Einstein was correct with his general relativity theory to explain accelerated expansion of the universe could also be explained by negative pressure working against gravity. The belief of Einstein to the static universe made him to think about negative pressure which will stop the attraction of the gravity. However, we know that we have non static universe, moreover we know that we have accelerated expansion. According to the above observational data analysis, it can be estimated the amount of the negative pressure in our universe, which we call it as dark energy. The simple question about the nature of the dark energy is still one of the intriguing questions and left free space for new speculations.\\
\\
After Einstein many Physicists have been investigating about gravitation in different contexts. Weyl [18] attempted to generalize the idea of geometrizing the gravitation and electromagnetism by applying different techniques and methods. He described both gravitation and electromagnetism geometrically by formulating a new kind of gauge theory involving metric tensor with an intrinsic geometrical significance.  With the concept of Einstein's general theory of relativity,  Lyra [19] suggested a modification  by introducing a gauge function into the structure less manifold which removes the non-integrability condition of the length of a vector under parallel transport. Such theories are commonly known as the modified theories of the gravitation or alternate theory of gravitation. Some important modified theories of gravitation are Brans-Dicke theory [20], Scalar-tensor theories[21], Vector-tensor theory[22], Weyl's theory [23], F(R) Gravity[24], Mimetic Gravity [25], Mimetic F(R)gravity [26], Lyra geometry [27] and many more. We can explain about the accelerating expansion of the universe in the context of these modified theories of gravitation. Out of these modified theories of gravitation, here we will discuss about the Lyra geometry.\\
\\
As we know that Lyra geometry is a modification of Riemanian geometry by introducing a gauge function into the structure less manifold which removes the non-integrability condition of the length of a vector under parallel transport. Lyra geometry along with constant gauge vector  $\phi_i$ will either play the role of cosmological constant or creation field ( equal to Hoyle's creation field [28-30]) which is discussed by Soleng [31]. Many researchers proposed different cosmological models in different context of Lyra geometry.  Sen [32] formulated a new scalar-tensor theory of gravitation based on Lyra geometry in which he found that static model with finite density in Lyra geometry is similar to the  Einstein's static model, but it exhibited red shift which is a significant difference of the model. Later, Sen and Dunn [33], Rosen [34], suggested that this theory was based on non-integrability of length transfer so that it had some unsatisfactory features and hence this theory did not gain general acceptance. Halford [35, 36] pointed out that in the normal general relativistic treatment the constant displacement vector field $\phi_{k}$ in Lyra's geometry plays the role of cosmological constant. Also, as in the Einstein's theory of relativity, the scalar-tensor treatment based on Lyra's geometry predicts the same effect, within observational limits, as far as the classical solar system test are concerned. any authors (Rahaman et al. [37, 38], Casana et al. [39, 40] , Mohanty et al. [41, 42] , Mahanta et al. [43], Asgar et al. [44], Mollah et al.[45], Mollah and Priyokumar [46]) attempted to solve Einstein's field equations in the framework of Lyra's geometry and successfully find their solutions under different circumstances.\\
\\
From the critical study, we know that space-time symmetry plays a vital role in the features of space-time that can be described as exhibiting some form of symmetry. The importance of symmetries in relativity and cosmology is to simplify Einstein's field equations and to provide a classification of the space time according to the structure of the corresponding Lie algebra. Some authors (Henriksen et al. [47] , Mohanty et al. [48] , Rao et al. [49] , Cahill et al. [50]) solved Einstein's field equations for an axially symmetric space-time by imposing certain conditions upon the scale factor of the space time together with some conditions on matter which represents such space-time and some restrictions on its physical properties. To explain the accelerated expansion of the universe many prominent researchers such as Shchigolev et al. [51] , Hova [52] , Ali et al. [53] , Megied et al. [54],  Khurshudyan  et al. [55, 56],  Saadat [57] , Darabi et al. [58] , Ziaie  et al. [59] , Pucheu  et al. [60] have investigated and proposed different cosmological models and ideas of the universe within the framework of Lyra's geometry and other theories of relativity in different context.\\
\\
It seems that, majority of  researchers in the field of Astrophysics, Cosmology and Particle Physics were doing their research in this area by considering only 4-dimensional case. But in this modern era, peoples are more interested to study higher dimensional case, since the solutions of Einstein field equations in higher dimensional space times are believed to have physical relevance possibly at extremely early times before the universe underwent the compactification transitions. We also found that by using a suitable scalar field we can show that the phase transitions on the early universe can give rise to such objects which are nothing but the topological knots in the vacuum expectation value of the scalar field and most of their energy is concentrated in a small region.  So it is necessary for us to study the cosmological problems by considering higher-dimensional space-time  to unify gravity with other interactions. In cosmology, particularly for study of the early stage of universe the present four dimensional stage of the universe might have been preceded by a multi-dimensional stage. Recently, Priyokumar and Mollah [61] studied about higher dimensional LRS Bianchi type-I cosmological model universe interacting with perfect fluid in Lyra geometry with different cases.  Many authors [62-70] have studied Bianchi type models in order to examine the role of certain anisotropic sources during the formation of the large-scale structures that we see in the present universe and for better understanding the small amount of observed anisotropy in the universe. Banerjee et al.,[71] have investigated Bianchi type-I cosmological models with viscous fluid in higher dimensional space time. Also Krori et al.[72] studied Bianchi type-I string cosmological model in higher dimensions.\\
\\
Motivated from the above, in this paper we discussed about five-dimensional spatially homogeneous and anisotropic locally rotationally symmetric (LRS) Bianchi type-I universe with time dependent deceleration parameters in  Lyra manifold considering different cases and find out the realistic solutions which are supporting to the present observational facts. We also found that the geometry itself of Lyra manifold behaves as a new source of dark energy which will be beneficial for further research work.\\
\\
This paper is organized as follows. In Sec. 2, we are presenting the formulation of the problem and the physical properties for the model are defined. In Sec. 3, considering deceleration parameter in four different forms, we have obtained six different cosmological models (in four cases together with three subcases of case-II) by solving the field equations obtained in the previous section. Some important physical parameters representing the models are also obtained in this section. In Sec. 4, Physical interpretations of all the model universes are discussed. In Sec. 5, conclusion and summary of our work is presented.

\section{Formulation of problem}

The Einstein's field equations based on Lyra geometry is proposed by Sen [32] and Sen and Dunn [33] in normal gauge may be written as
\begin{equation}
R_{ij}-\frac{1}{2}g_{ij}R+\frac{3}{2}\phi_{i}\phi_{j}-\frac{3}{4}g_{ij}\phi^k\phi_k=-T_{ij}
\end{equation}
where $\phi_i$ is the displacement vector and other symbols have their usual meanings as in the Riemannian geometry. The displacement vector $\phi_i$ is taken in the form

\begin{equation}
\phi_i=(0,0,0,0,\beta(t))
\end{equation}
\\
In this Paper we consider the five dimensional LRS Bianchi type-I axially symmetric space time in the form
\begin{equation}
ds^{2}=A^{2}(dx^{2}+dy^{2}+dz^{2})+B^{2}d\psi^{2}-dt^{2}
\end{equation}
where A and  B are functions of cosmic time $'t'$ only, where the fifth coordinate is taken to be space-like. Here the spatial curvature has been taken as zero (see [73]). \\
\\
The energy momentum tensor $T_{ij}$ for the perfect fluid is given by
\begin{equation}
T_{ij}=(\rho + p)u_{i}u_{j}+pg_{ij}
\end{equation}
where, $\rho$  is the energy density , $p$ is the fluid pressure and $u^{i}$ is the five velocity vector given by $u^{i}=(0,0,0,0,1)$ satisfying
\begin{equation}
g_{ij}u^{i}u^{j}=u^{i}u_{i}=-1
\end{equation}
\\
The physical quantities like Spatial Volume $V$ , Hubble's Parameter $H$ , Expansion scalar $\theta$ , Shear scalar $\sigma$ , mean anisotropy parameter $\Delta$ play a vital role for a cosmological model. These parameters for the axially symmetric LRS Bianchi type-I metric (3) are defined as

\begin{equation}
V=R^{4}(t)=A^{3}B
\end{equation}

\begin{equation}
H=\frac{\dot{R}}{R}=\frac{1}{4}\left(3\frac{\dot{A}}{A}+\frac{\dot{B}}{B}\right)
\end{equation}

\begin{equation}
\theta=4H=3\frac{\dot{A}}{A}+\frac{\dot{B}}{B}
\end{equation}

\begin{equation}
\sigma^{2}=\frac{1}{2}\left(\sum^{4}_{i=1}H_{i}^{2}-4H^{2}\right)
\end{equation}

\begin{equation}
\Delta=\frac{1}{4}\sum^{4}_{i=1}\left(\frac{H_{i}-H}{H}\right)^{2}
\end{equation}
\\
where $H_{i}$ ; $i=1,2,3,4$ represent the directional Hubble's parameters in $x, y, z, \psi$ directions respectively and $\Delta=0$ corresponds to isotropic expansion.\\
\\
Using the comoving coordinate system, the field equations (1) for the metric (3) with the help of equation (4) yield

\begin{equation}
3\frac{\dot{A}^2}{A^2}+3\frac{\dot{A}\dot{B}}{AB}-\frac{3}{4}\beta^2=-\rho
\end{equation}

\begin{equation}
2\frac{\ddot{A}}{A}+\frac{\ddot{B}}{B}+\frac{\dot{A}^2}{A^2}+2\frac{\dot{A}\dot{B}}{AB}+\frac{3}{4}\beta^2=-p
\end{equation}

\begin{equation}
3\frac{\ddot{A}}{A}+3\frac{\dot{A}^2}{A^2}+\frac{3}{4}\beta^2=-p
\end{equation}
\\
where the overhead dots $(.)$ denote the derivatives with respect to time $t$.\\
\\
Again the energy conservation equation $T^{ij}_{;j}=0$ gives us

\begin{equation}
\dot{\rho}+(\rho+p)\left(3\frac{\dot{A}}{A}+\frac{\dot{B}}{B}\right)=0
\end{equation}
\\
The field equations (11)-(13) are a system of three highly nonlinear differential equations with five unknown parameters $A$, $B$, $\rho$, $p$ and $\beta$. Therefore in order to obtain deterministic solution of the above system of equations, it is required to assume two more additional physical conditions/mathematical relations  involving these unknowns. Hence we assume\\
\\
(a) The expansion scalar $\theta$ is proportional to the shear scalar $(\sigma)$ (see [74]) which leads to
        \begin{equation}
        A=B^{n}
        \end{equation}
where $n\neq1$ is a constant and \\
\\
(b) Another additional condition is taken as the any one of the following time dependent deceleration parameter \\

(i) Average scale factor as an integrating function of time (see [75]) as

        \begin{equation}
        R=\left(t^{a}e^{t}\right)^{\frac{1}{b}}
        \end{equation}
\\
where $a$ and $b$ are positive constants in which scale factor has two factors: one factor behaving like exponential expansion and the other factor behaving like power law expansion. While the power law behaviour dominate the cosmic dynamics in early phase of cosmic evolution, the exponential factor dominates at late phase. When $a = 0$, the exponential law is recovered and for $1/b = 0$, the scale factor reduces to the power law as suggested by Mishra [76] and Akarsu et al. [77]. Here we are very much interested in a transient universe with early deceleration and late acceleration so that at early time $q$ can be positive whereas at late time $q$ assumes a negative value in conformity with the recent observational data.\\
\\
    Therefore, the value of deceleration parameter (DP) $q$ becomes a function of time $t$ as follows \\

    \[q=-\frac{R\ddot{R}}{\dot{R}^{2}}=\frac{ab}{(a+b)^{2}}-1\]

(ii) Variable deceleration parameter

    \[q=-\frac{R\ddot{R}}{\dot{R}^{2}}=D\]

     where $R$ is the average scale factor and $D$ is a variable.\\

(iii) Linearly varying deceleration parameter

        \[q=-\frac{R\ddot{R}}{\dot{R}^{2}}=-kt+m-1\]

     where $R$ is the average scale factor, $k\geq0,  m\geq0$ are constants. \\

(iv) Special form of deceleration parameter

        \[q=-\frac{R\ddot{R}}{\dot{R}^{2}}=-1+\frac{c}{1+R^{c}}\]

     where $R$ is the average scale factor and $c>0$ is a constant. \\
\\
The above four  conditions will respectively give us the four different model universes as follows-
\\

\section{Solution of field equations in four different cases (of Deceleration Parameter)}

\textbf{Case - I : Models with Time Dependent Deceleration Parameter}\\
\\
In this case we consider the average scale factor as an integrating function of time as given in Eq. (16) as

 \[R=\left(t^{a}e^{t}\right)^{\frac{1}{b}}\]
\\
where $a$ and $b$ are positive constants.
\\
From the above value of the average scale factor $R$, the value of deceleration parameter $q$ can be obtained as
\\
\begin{equation}
q=-\frac{R\ddot{R}}{\dot{R}^{2}}=\frac{ab}{(a+t)^{2}}-1
\end{equation}
\\
This law (17) gives us a time-dependent deceleration parameter which describes that initially our universe was decelerating but as the the time progresses our universe becomes an accelerating. That is, our universe is in transition from early decelerating phase to the present accelerating phase.\\
\\
From Eq. (6) we have
\begin{equation}
R(t)=\left(A^{3}B\right)^{\frac{1}{4}}
\end{equation}
\\
The metric (1) can be determined completely by this average scale factor $R(t)$ given by Eq. (18).\\
\\
From Eqs. (15), (16) and (18) we have

\begin{equation}
A=\left(t^{a}e^{t}\right)^{\frac{4n}{b(3n+1)}}
\end{equation}
and
\begin{equation}
B=\left(t^{a}e^{t}\right)^{\frac{4}{b(3n+1)}}
\end{equation}
\\
By using Eqs. (19) and (20) the metric (3) can be written as

\begin{equation}
ds^{2}=\left(t^{a}e^{t}\right)^{\frac{8n}{b(3n+1)}}(dx^{2}+dy^{2}+dz^{2})+\left(t^{a}e^{t}\right)^{\frac{8}{b(3n+1)}}d\psi^{2}-dt^{2}
\end{equation}
Equation (21) represents a Bianchi type-I cosmological model universe in Lyra geometry with time dependent deceleration parameter.\\
\\
\\
\textbf{Some Physical Properties of the Model (21) with Time Dependent Deceleration Parameter}\\
\\
The energy conservation equation (14) leads to

\[\frac{3}{2}\beta\dot{\beta}+\frac{3}{2}\beta^{2}\left(3\frac{\dot{A}}{A}+\frac{\dot{B}}{B}\right)=0\]

Now since the displacement vector $\beta\neq0$, so we have

\begin{equation}
\dot{\beta}+\beta\left(3\frac{\dot{A}}{A}+\frac{\dot{B}}{B}\right)=0
\end{equation}

From Eq. (22), the displacement vector $\beta$ is obtained as

\begin{equation}
\beta=\left(k_{0}t^{a}e^{t}\right)^{-\frac{4}{b}}
\end{equation}
\\
Energy density $\rho$ and pressure $p$ can be obtained from (11) and (13) as follows:

\begin{equation}
\rho=-\frac{48n(n+1)}{b^{2}(3n+1)^{2}}\left(\frac{a}{t}+1\right)^{2}+\frac{3}{4}\left(k_{0}t^{a}e^{t}\right)^{-\frac{8}{b}}
\end{equation}

\begin{equation}
p=-\frac{96n^{2}}{b^{2}(3n+1)^{2}}\left(\frac{a}{t}+1\right)^{2}+\frac{12an}{b(3n+1)^{2}}-\frac{3}{4}\left(k_{0}t^{a}e^{t}\right)^{-\frac{8}{b}}
\end{equation}
\\
Equations (6)-(10), respectively, gives us the Spatial Volume $V$ , Hubble's Parameter $H$ , Expansion scalar $\theta$ , Shear scalar $\sigma$ , mean anisotropy parameter $\Delta$ as follows:

\begin{equation}
V=\left(t^{a}e^{t}\right)^{\frac{4}{b}}
\end{equation}

\begin{equation}
H=\frac{\dot{R}}{R}=\frac{1}{b}\left(\frac{a}{t}+1\right)
\end{equation}

\begin{equation}
\theta=\frac{4}{b}\left(\frac{a}{t}+1\right)
\end{equation}

\begin{equation}
\sigma^{2}=\frac{6(n-1)^{2}}{b^{2}(3n+1)^{2}}\left(\frac{a}{t}+1\right)^{2}
\end{equation}

\begin{equation}
\Delta=\frac{3(n-1)^{2}}{(3n+1)^{2}}=constant ~(\neq0 ~~ for ~~ n\neq1)
\end{equation}
\
From Eqs. (28) and (29) we have
\[\frac{\sigma^{2}}{\theta^{2}}=\frac{3(n-1)^{2}}{8(3n+1)^{2}}~,~which ~ is~a~ constant~ and~is~ independent~ of~ time ~ t\]
Hence
\begin{equation}
\lim_{t\to \infty}\frac{\sigma^{2}}{\theta^{2}}=\frac{3(n-1)^{2}}{8(3n+1)^{2}} ~\neq0 ~ for ~ n\neq1
\end{equation}\\
\\
The variation of some parameters with respect to time for this case are shown in Figs. 1-3 \\

\includegraphics{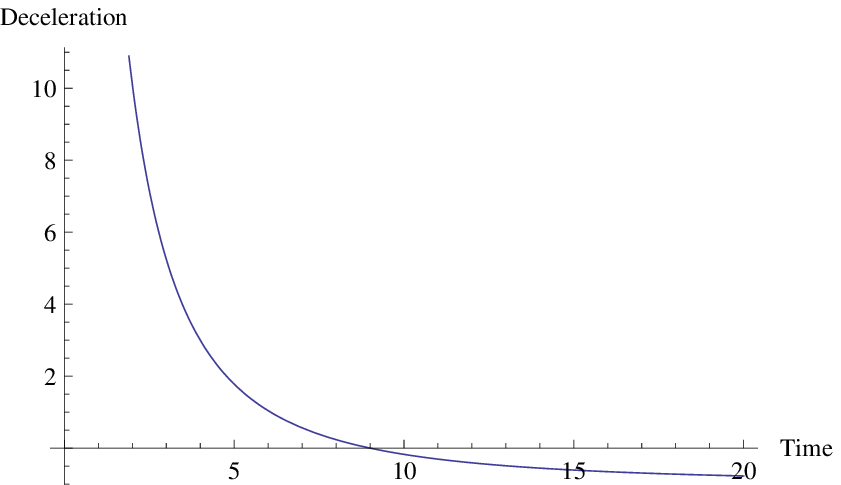}
\\
\textbf{Fig. 1 : The plot of Deceleration Parameter q vs. Time t. Here, $a= k_0=1, b=100$ and $n=2$.}\\
\\
\includegraphics{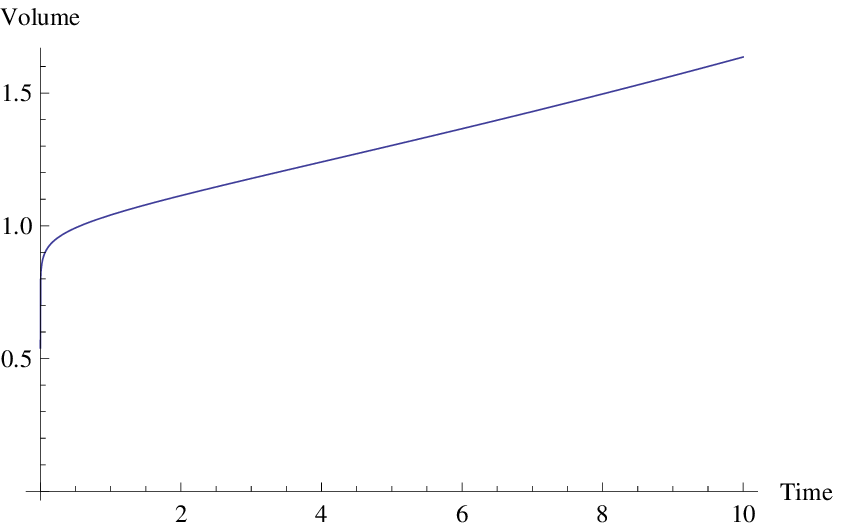}
\\
\textbf{Fig. 2 : The plot of Volume V vs. Time t. Here, $a= k_0=1, b=100$ and $n=2$.}\\
\\

\includegraphics{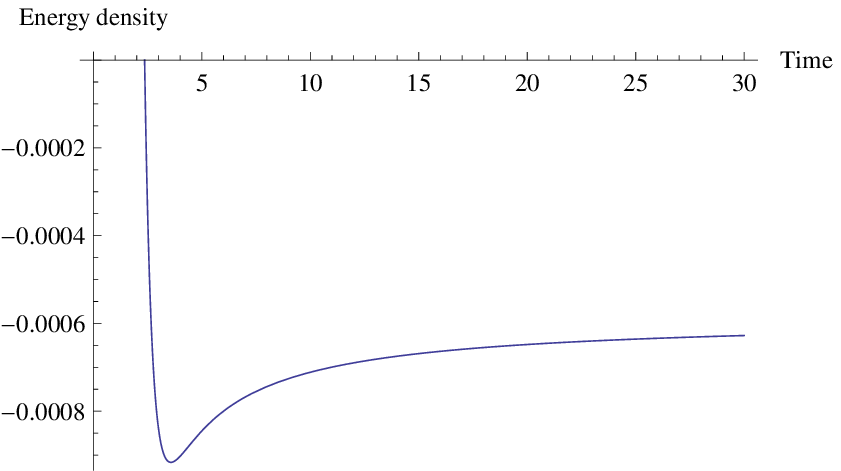}
\\
\textbf{Fig. 3 : The plot of Energy density $\rho$ vs. Time t. Here, $a= k_0=1, b=100$ and $n=2$.}\\
\\
\\
\textbf{Case - II : Models with Time Dependent Deceleration Parameter}\\
\\
Let us consider that the deceleration parameter $q$ to be a variable parameter (see [78]) as

\begin{equation}
q=-\frac{R\ddot{R}}{\dot{R}^{2}}=D ~~ (variable)
\end{equation}
\\
where $R$ is the average scale factor.\\
\\
Equation (32) can be written as
\begin{equation}
\frac{\ddot{R}}{\dot{R}}+D\frac{\dot{R}^{2}}{R^{2}}=0
\end{equation}
In order to solve equation (33), we may consider $D=D(R)$ . Now since R is a function of time, so we can assume $D=D(t)=D\left(R(t)\right)$ and it is possible only when there is one to one correspondence between $R$ and $t$ . Again since both $R$ and $t$ are increasing function so this is possible only we can avoid singularity like big bang or big rip. \\
\\
Therefore if we assume $D=D(R)$ then the general solution of equation (33) is obtained as
\begin{equation}
\int{e^{\int\frac{B}{R}\,dR}}\,dR=t+c_{0}
\end{equation}
where $c_{0}$ is an integrating constant. Without loss of generality, we may choose $c_{0}=0$\\
\\
When $c_{0}=0$ then in order to solve (34), we have to choose $\int\frac{B}{R}\,dR$ in such a way that (34) becomes integrable. Hence we consider
\begin{equation}
\int\frac{B}{R}\,dR=\log L(R)
\end{equation}
where $L(R)$ is a function of $R$ only and it does not affect the nature of generality of solution.\\
\\
From Eqs. (34) and (35) we have
\begin{equation}
\int L(R)\,dR=t
\end{equation}
Since the choice of function $L(R)$ in equation (36) is quite arbitrary and we are interested in obtaining the models of the universe which are not only physically viable but also consistent with observations, so we have chosen the function $L(R)$ in three different ways
\\
\[L(R)=\frac{1}{k_{1}R}\]
where $k_{1}$ is an arbitrary constant.

\[L(R)=\frac{1}{2k_{3}\sqrt{R+k_{4}}}\]
where $k_{3}$ and $k_{4}$ are arbitrary constants.

\[L(R)=\frac{nR^{n-1}}{k_{5}\sqrt{1+R^{2n}}}\]
where $k_{5}$ is an arbitrary constant.\\
\\
\textbf{Subcase-1 of Case II :} In the First case, the function $L(R)$ is chosen as
\\
\begin{equation}
L(R)=\frac{1}{k_{1}R}
\end{equation}
where $k_{1}$ is an arbitrary constant.\\
\\
For this case, integrating Eq. (36), we may obtain the exact solution as
\begin{equation}
R=k_{2}e^{k_{1}t}
\end{equation}
Where $k_{2}$ is an integrating constant.
\\
From Eqs. (6), (15) and (38), we have
\begin{equation}
A=k_{2}^{\frac{4n}{3n+1}}e^{\frac{4nk_{1}t}{3n+1}}
\end{equation}
\begin{equation}
B=k_{2}^{\frac{4}{3n+1}}e^{\frac{4k_{1}t}{3n+1}}
\end{equation}
\\
Using the values of $A$ and $B$ from Eqs. (39) and (40), the metric (3) can be written as

\begin{equation}
ds^{2}=k_{2}^{\frac{8n}{3n+1}}e^{\frac{8nk_{1}t}{3n+1}}(dx^{2}+dy^{2}+dz^{2})+k_{2}^{\frac{8}{3n+1}}e^{\frac{8k_{1}t}{3n+1}}d\psi^{2}-dt^{2}
\end{equation}
Equation (41) represents a Bianchi type-I cosmological model universe in Lyra geometry with variable deceleration parameter.\\
\\
\textbf{Some Physical Properties of the Model (41) with variable Deceleration Parameter}\\
\\
Equations (6)-(10), respectively, give us the Spatial Volume $V$ , Hubble's Parameter $H$ , Expansion scalar $\theta$ , Shear scalar $\sigma$ , mean anisotropy parameter $\Delta$ as follows:

\begin{equation}
V=k_{2}^{4}e^{4k_{1}t}
\end{equation}

\begin{equation}
H=k_{1}=constant
\end{equation}

\begin{equation}
\theta=4k_{1}=constant
\end{equation}

\begin{equation}
\sigma^{2}=\frac{6(n-1)^{2}k_{1}^{2}}{(3n+1)^{2}}
\end{equation}

\begin{equation}
\Delta=\frac{3(n-1)^{2}}{(3n+1)^{2}}
\end{equation}
\\
From Eqs. (44) and (45) we have
\[\frac{\sigma^{2}}{\theta^{2}}=\frac{3(n-1)^{2}}{8(3n+1)^{2}}~,~which ~ is~a~constant~and~is~ independent~ of~ time ~ t\]
\\
Hence
\begin{equation}
\lim_{t\to \infty}\frac{\sigma^{2}}{\theta^{2}}=\frac{3(n-1)^{2}}{8(3n+1)^{2}} ~\neq0 ~ for ~ n\neq1
\end{equation}\\
\\
Again from (22), (11) and (13), the displacement vector $\beta$, the energy density $\rho$ and the fluid pressure $p$ are obtained as follows:
\begin{equation}
\beta=l_{2}e^{-4k_{1}t}
\end{equation}

\begin{equation}
\rho=-\frac{48k_{1}^{2}n(n+1)}{(3n+1)^{2}}+\frac{3}{4}l_{2}e^{-4k_{1}t}
\end{equation}

\begin{equation}
p=-\frac{96n^{2}k_{1}^{2}}{(3n+1)^{2}}-\frac{3}{4}l_{2}e^{-4k_{1}t}
\end{equation}
\\
Where $l_{2}$ is a constant.\\
\\
Also the deceleration parameter $q$ is obtained as follows:
\begin{equation}
q=-1
\end{equation}\\
\\
The variation of some parameters with respect to time for this sub case 1 are shown in Figs. 4-8 \\

\includegraphics{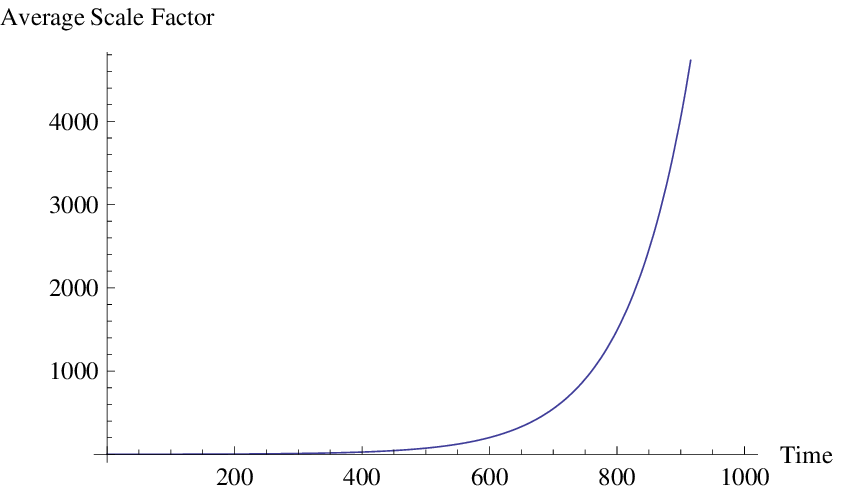}
\\
\\
\textbf{Fig. 4 : The plot of Average Scale factor $R(t)$ vs. Time $t$. Here, $k_{1}=0.01$, $k_{2}=0.5$ and $n=2$}\\
\\
\\
\includegraphics{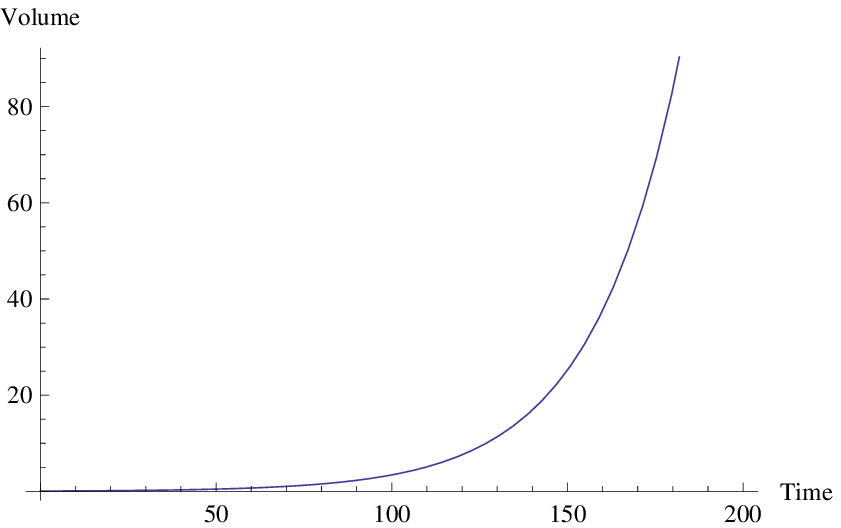}
\\
\\
\textbf{Fig. 5 : The plot of Volume $V$ vs. Time $t$. Here, $k_{1}=0.01$, $k_{2}=0.5$ and $n=2$}\\
\\
\\
\includegraphics{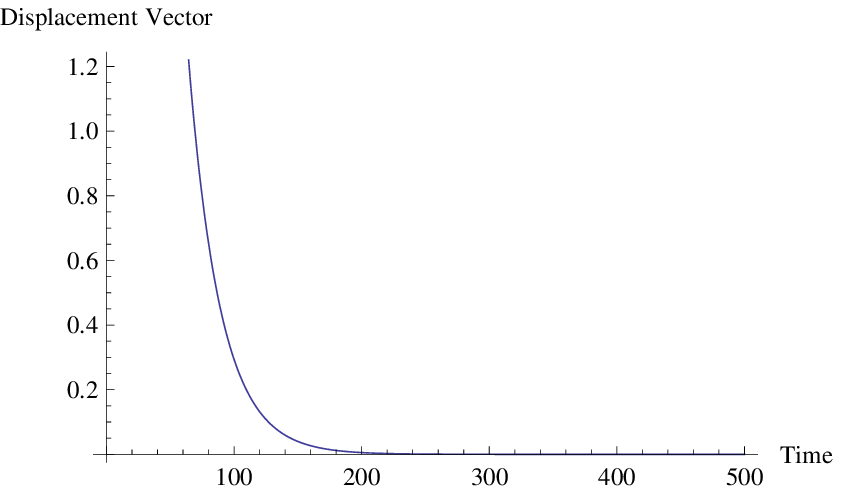}
\\
\textbf{Fig. 6 : The plot of Displacement vector $\beta$ vs. Time $t$. Here, $k_{1}=0.01$, $k_{2}=0.5$, $n=2$ and $l_{2}=16$}\\
\\
\\
\includegraphics{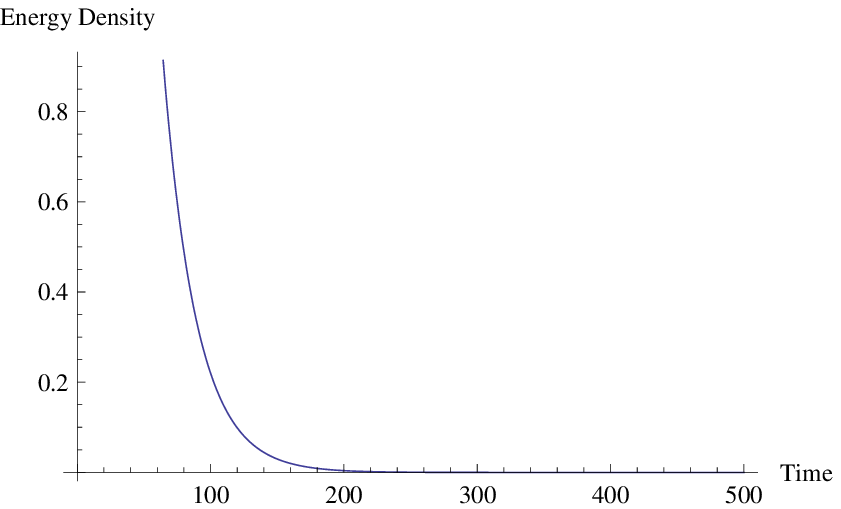}
\\
\textbf{Fig. 7 : The plot of Energy density $\rho$ vs. Time $t$. Here, $k_{1}=0.01$, $k_{2}=0.5$, $n=2$ and $l_{2}=16$}\\
\\
\includegraphics{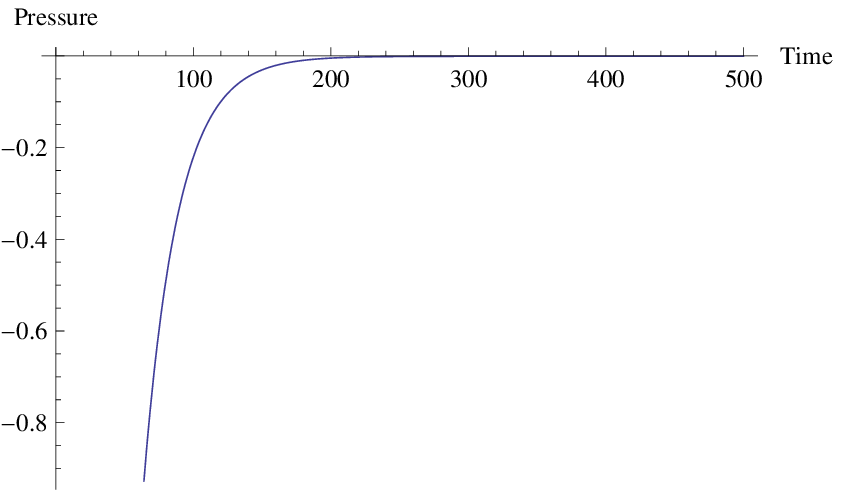}
\\
\textbf{Fig. 8 : The plot of Pressure $p$ vs. Time $t$. Here, $k_{1}=0.01$, $k_{2}=0.5$, $n=2$ and $l_{2}=16$}\\
\\
\\
\textbf{Subcase-2 of Case II :} In the second case we choose $L(R)$ as
\\
\begin{equation}
L(R)=\frac{1}{2k_{3}\sqrt{R+k_{4}}}
\end{equation}
where $k_{3}$ and $k_{4}$ are arbitrary constants.
\\
Therefore from Eq. (36) we get
\begin{equation}
R=\alpha_{1}t^{2}+\alpha_{2}t+\alpha_{3}
\end{equation}
where $\alpha_{1}$, $\alpha_{2}$ and $\alpha_{3}$ are arbitrary constants.
\\
This value of $R$ gives us the values of $A$ and $B$, from Eqs. (6) and (15), as follows
\begin{equation}
A=(\alpha_{1}t^{2}+\alpha_{2}t+\alpha_{3})^{\frac{4n}{3n+1}}
\end{equation}
\begin{equation}
B=(\alpha_{1}t^{2}+\alpha_{2}t+\alpha_{3})^{\frac{4}{3n+1}}
\end{equation}
\\
Using the values of $A$ and $B$ from Eqs. (54) and (55) in the metric (3), we have

\begin{equation}
ds^{2}=(\alpha_{1}t^{2}+\alpha_{2}t+\alpha_{3})^{\frac{8n}{3n+1}}(dx^{2}+dy^{2}+dz^{2})+(\alpha_{1}t^{2}+\alpha_{2}t+\alpha_{3})^{\frac{8}{3n+1}}d\psi^{2}-dt^{2}
\end{equation}
\\
Equation (56) give us also another form of Bianchi type-I cosmological model universe in Lyra geometry with variable deceleration parameter.\\
\\
\textbf{Some Physical Properties of the Model (56) with variable Deceleration Parameter}\\
\\
Equations (6)-(10), respectively, give us the Spatial Volume $V$ , Hubble's Parameter $H$ , Expansion scalar $\theta$ , Shear scalar $\sigma$ , mean anisotropy parameter $\Delta$ as follows:

\begin{equation}
V=(\alpha_{1}t^{2}+\alpha_{2}t+\alpha_{3})^{4}
\end{equation}

\begin{equation}
H=\frac{2\alpha_{1}t+\alpha_{2}}{\alpha_{1}t^{2}+\alpha_{2}t+\alpha_{3}}
\end{equation}

\begin{equation}
\theta=\frac{4(2\alpha_{1}t+\alpha_{2})}{\alpha_{1}t^{2}+\alpha_{2}t+\alpha_{3}}
\end{equation}

\begin{equation}
\sigma^{2}=\frac{6(n-1)^{2}}{(3n+1)^{2}}\frac{(2\alpha_{1}t+\alpha_{2})^{2}}{(\alpha_{1}t^{2}+\alpha_{2}t+\alpha_{3})^{2}}
\end{equation}

\begin{equation}
\Delta=\frac{3(n-1)^{2}}{(3n+1)^{2}}= ~ constant ~(\neq0 ~~ for ~~ n\neq1)
\end{equation}
\\
From Eqs. (59) and (60) we have

\[\frac{\sigma^{2}}{\theta^{2}}=\frac{3(n-1)^{2}}{8(3n+1)^{2}}~,~which ~ is~a~ constant~ and~is~ independent~ of~ time ~ t\]
\\
Hence
\begin{equation}
\lim_{t\to \infty}\frac{\sigma^{2}}{\theta^{2}}=\frac{3(n-1)^{2}}{8(3n+1)^{2}} ~\neq0 ~ for ~ n\neq1
\end{equation}\\
\\
Again from (22), (11) and (13), the displacement vector $\beta$, the energy density $\rho$ and the fluid pressure $p$ are obtained as follows:
\begin{equation}
\beta=\frac{\alpha_{4}}{(\alpha_{1}t^{2}+\alpha_{2}t+\alpha_{3})^{4}}
\end{equation}

\begin{equation}
\rho=-\frac{48(n+1)}{(3n+1)^{2}}\frac{(2\alpha_{1}t+\alpha_{2})^{2}}{(\alpha_{1}t^{2}+\alpha_{2}t+\alpha_{3})^{2}}+
        \frac{3}{4}\frac{\alpha_{4}^{2}}{(\alpha_{1}t^{2}+\alpha_{2}t+\alpha_{3})^{8}}
\end{equation}

\begin{equation}
\begin{split}
p=&-\frac{12n(5n-1)}{(3n+1)^{2}}\frac{(2\alpha_{1}t+\alpha_{2})^{2}}{(\alpha_{1}t^{2}+\alpha_{2}t+\alpha_{3})^{2}}-
    \frac{24n}{3n+1}\frac{\alpha_{1}}{(\alpha_{1}t^{2}+\alpha_{2}t+\alpha_{3})^{2}}\\
        &-\frac{3}{4}\frac{\alpha_{4}^{2}}{(\alpha_{1}t^{2}+\alpha_{2}t+\alpha_{3})^{8}}
\end{split}
\end{equation}
\\
Also the deceleration parameter $q$ is obtained as follows:
\begin{equation}
q=-\frac{2\alpha_{1}(\alpha_{1}t^{2}+\alpha_{2}t+\alpha_{3})}{(2\alpha_{1}t+\alpha_{2})^{2}}
\end{equation}\\
\\
The variation of some parameters with respect to time for this sub case 2 are shown in Figs. 9-11 \\

\includegraphics{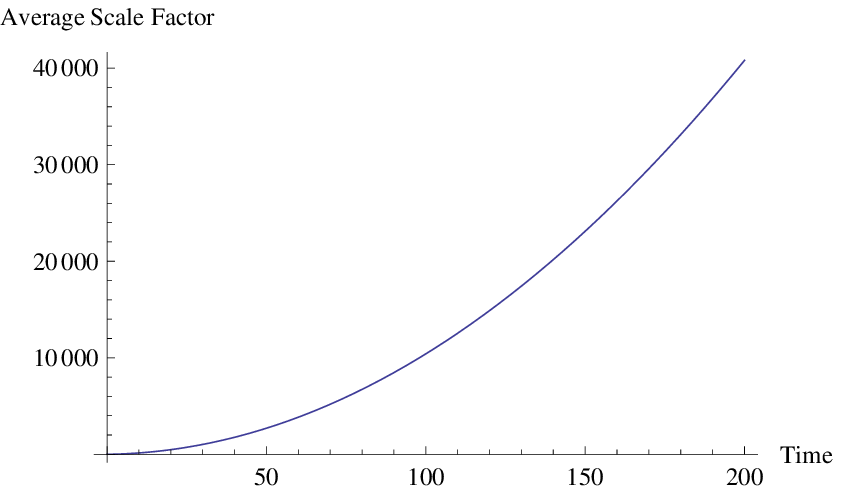}
\\
\\
\textbf{Fig. 9 : The plot of Average Scale Factor $R(t)$ vs. Time $t$. Here, $\alpha_{1}=1$, $\alpha_{2}=4$, $\alpha_{3}=3$, $\alpha_{4}=1$ and $n=2$}\\
\\
\\
\includegraphics{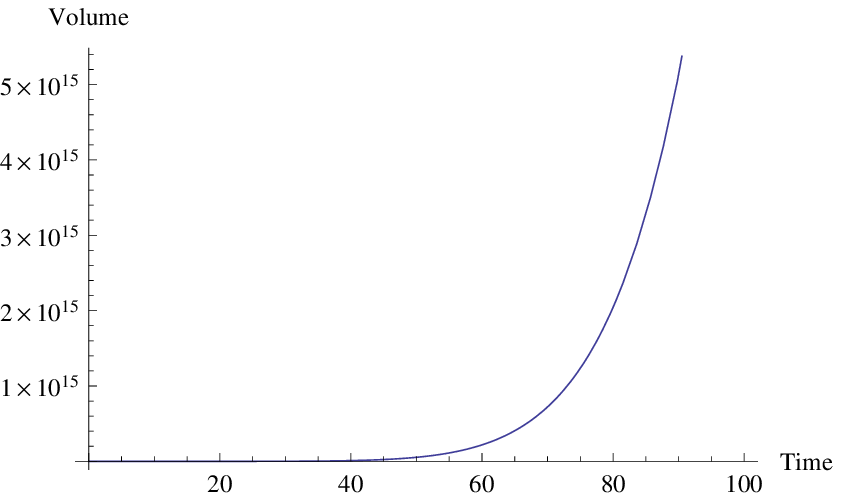}
\\
\\
\textbf{Fig. 10 : The plot of Volume $V$ vs. Time $t$. Here, $\alpha_{1}=1$, $\alpha_{2}=4$, $\alpha_{3}=3$, $\alpha_{4}=1$ and $n=2$}\\
\\
\\
\includegraphics{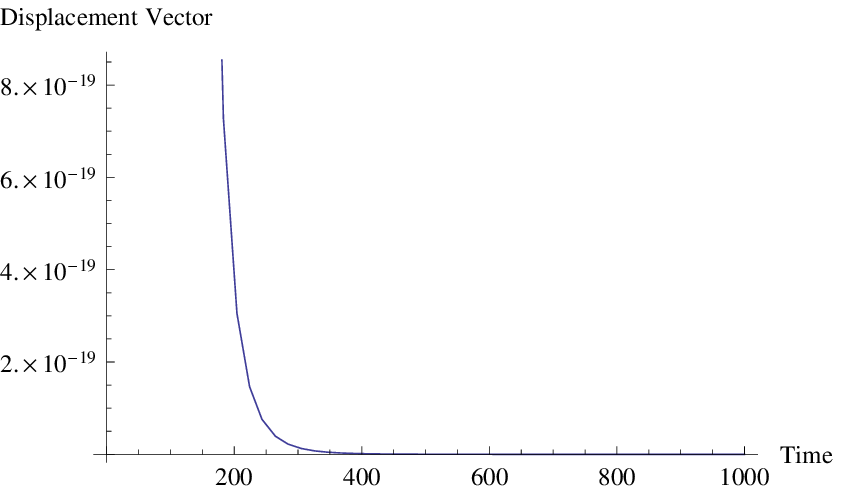}
\\
\\
\textbf{Fig. 11(a): The plot of Displacement Vector $\beta(t)$ vs. Time $t$. Here, $\alpha_{1}=1$, $\alpha_{2}=4$, $\alpha_{3}=3$, $\alpha_{4}=1$ and $n=2$}\\
\\
\\
\includegraphics{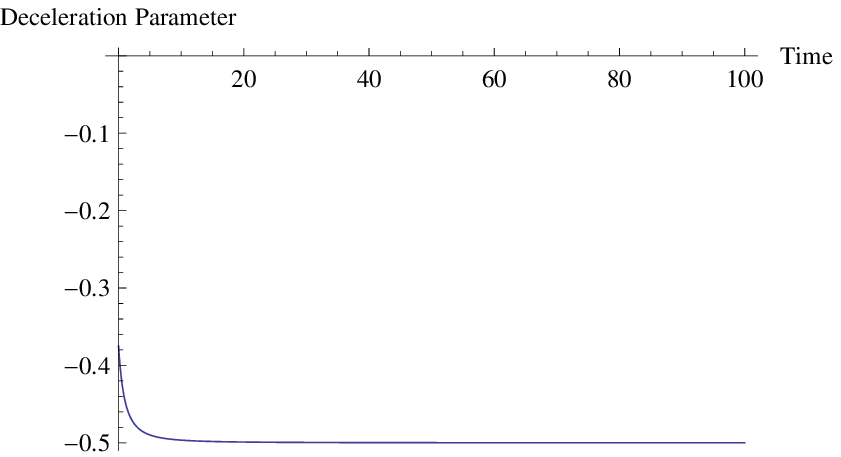}
\\
\\
\textbf{Fig. 11(b): The plot of Deceleration Parameter $q$ vs. Time $t$. Here, $\alpha_{1}=1$, $\alpha_{2}=4$, $\alpha_{3}=3$, $\alpha_{4}=1$ and $n=2$}\\
\\
\\
\textbf{Subcase-3 of Case II :} Also in the third case we choose $L(R)$ as follows:
\begin{equation}
L(R)=\frac{nR^{n-1}}{k_{5}\sqrt{1+R^{2n}}}
\end{equation}
where $k_{5}$ is an arbitrary constant.\\
\\
Using (67) in (36), we have
\begin{equation}
R=[\sinh(k_{5}t)]^{\frac{1}{n}}
\end{equation}
From Eqs. (6), (15) and (68), we have
\begin{equation}
A=[\sinh(k_{5}t)]^{\frac{4}{3n+1}}
\end{equation}

\begin{equation}
B=[\sinh(k_{5}t)]^{\frac{4}{n(3n+1)}}
\end{equation}
\\
By the use of Eqs. (69) and (70), the metric (3) takes the form

\begin{equation}
ds^{2}=[\sinh(k_{5}t)]^{\frac{8}{3n+1}}(dx^{2}+dy^{2}+dz^{2})+[\sinh(k_{5}t)]^{\frac{4}{n(3n+1)}}d\psi^{2}-dt^{2}
\end{equation}
Equation (71) represents a Bianchi type-I cosmological model universe in Lyra geometry with variable deceleration parameter.\\
\\
\textbf{Some Physical Properties of the Model (71) with variable Deceleration Parameter}\\
\\
Equations (6)-(10), respectively, give us the Spatial Volume $V$ , Hubble's Parameter $H$ , Expansion scalar $\theta$ , Shear scalar $\sigma$ , mean anisotropy parameter $\Delta$ as follows:

\begin{equation}
V=[\sinh(k_{5}t)]^{\frac{4}{n}}
\end{equation}

\begin{equation}
H=\frac{\dot{R}}{R}=\frac{k_{5}}{n}\coth(k_{5}t)
\end{equation}

\begin{equation}
\theta=\frac{4k_{5}}{n}\coth(k_{5}t)
\end{equation}

\begin{equation}
\sigma^{2}=\frac{6(n-1)^{2}}{n^{2}(3n+1)^{2}}k_{5}^{2}\coth^{2}(k_{5}t)
\end{equation}

\begin{equation}
\Delta=\frac{3(n-1)^{2}}{(3n+1)^{2}}= ~ constant ~(\neq0 ~~ for ~~ n\neq1)
\end{equation}
\\
From Eqs. (74) and (75) we have
\[\frac{\sigma^{2}}{\theta^{2}}=\frac{3(n-1)^{2}}{8(3n+1)^{2}}~,~which ~ is~a~ constant~ and~is~ independent~ of~ time ~ t.\]
\
Hence
\begin{equation}
\lim_{t\to \infty}\frac{\sigma^{2}}{\theta^{2}}=\frac{3(n-1)^{2}}{8(3n+1)^{2}} ~\neq0 ~ for ~ n\neq1
\end{equation}\\
Again from (22), (11) and (13), the displacement vector $\beta$, the energy density $\rho$ and the fluid pressure $p$ are obtained as follows:
\begin{equation}
\beta=[\sinh(k_{5}t)]^{-\frac{4}{n}}
\end{equation}
\begin{equation}
\rho=-\frac{48(n+1)}{n(3n+1)}k_{5}^{2}\coth^{2}(k_{5}t)+\frac{3}{4}[\sinh(k_{5}t)]^{-\frac{8}{n}}
\end{equation}
\begin{equation}
p=-\frac{96k_{5}^{2}}{(3n+1)^{2}}\coth^{2}(k_{5}t)+\frac{12k_{5}^{2}}{(3n+1)}[\coth^{2}(k_{5}t)-1]-\frac{3}{4}[\sinh(k_{5}t)]^{-\frac{8}{n}}
\end{equation}
Also the deceleration parameter $q$ is obtained as follows:
\begin{equation}
q=n[1-\tanh^{2}(k_{5}t)]-1
\end{equation}\\
\\
The variation of some parameters with respect to time for this sub case 3 are shown in Figs. 12-14 \\
\includegraphics{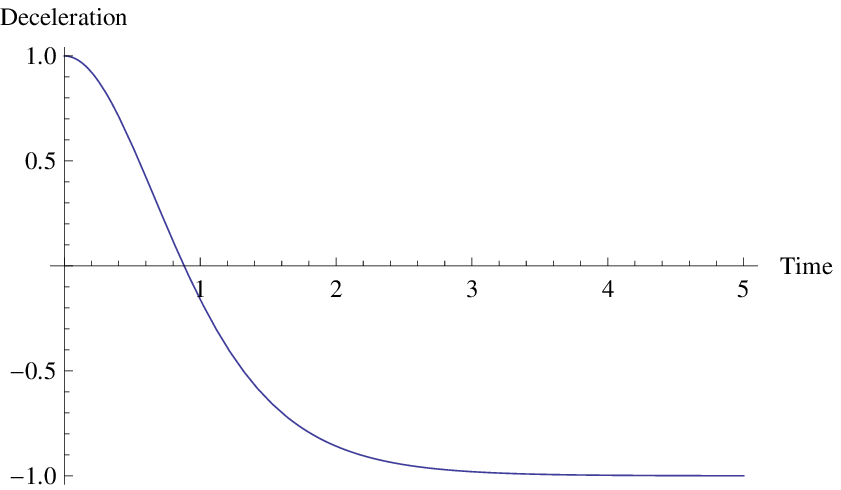}
\\
\\
\textbf{Fig. 12 : The plot of Deceleration Parameter q vs. Time t. Here $k_{5}=1$ and $n=2$}\\
\\
\includegraphics{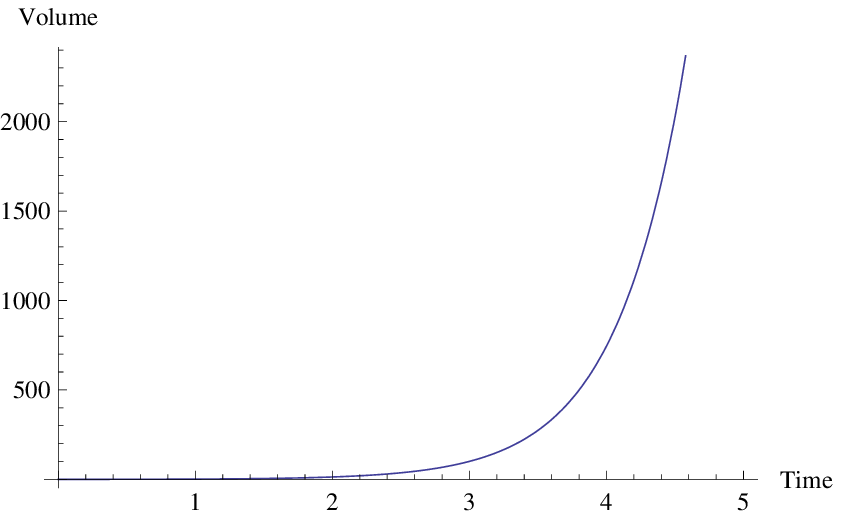}
\\
\\
\textbf{Fig. 13 : The plot of Volume V vs. Time t. Here $k_{5}=1$ and $n=2$}\\
\\
\includegraphics{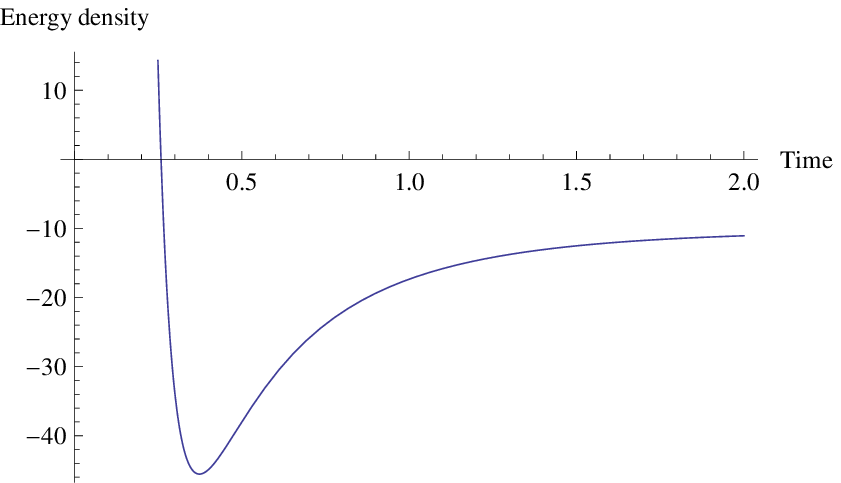}
\\
\\
\textbf{Fig. 14 : The plot of Energy density $\rho$ vs. Time t. Here $k_{5}=1$ and $n=2$}\\
\\
\\
\textbf{Case - III : Models with Linearly Varying Deceleration Parameter}\\
\\
In this case we consider the linearly varying deceleration parameter (see [79]) as follows:
\begin{equation}
q=-\frac{R\ddot{R}}{\dot{R}^{2}}=-kt+m-1
\end{equation}
where $R$ is the average scale factor, $k\geq0 ~,~ m\geq0$ are constants.\\
\\
For $k=0$ , the Eq. (52) reduces to
\begin{equation}
q=m-1
\end{equation}
which is a constant and it corresponds to cosmological model with constant deceleration parameter.\\
\\
Therefore, the cosmological models that are obtained via constant deceleration parameter may be generalized by using this law.\\
\\
By solving  Eq. (82), We obtain three different forms of the average scale factors as
\begin{equation}
R=R_{0}e^{\frac{2}{\sqrt{m^{2}-kc_{0}}}\tanh^{-1}\left(\frac{kt-m}{m^{2}-kc_{0}}\right)} ~~; ~~ for ~~ k>0 ~ and ~ m\geq0
\end{equation}

\begin{equation}
R=R_{1}\left(mt+c_{1}\right) ~~~~~~~~~~~~~~~~~~~~; ~~ for ~~ k=0 ~ and ~ m>0
\end{equation}

\begin{equation}
R=R_{2}e^{c_{2}t} ~~~~~~~~~~~~~~~~~~~~~~~~~~~~~; ~~ for ~~ k=0 ~ and ~ m=0
\end{equation}
where $R_{0}$ , $R_{1}$ , $R_{2}$ , $c_{0}$ , $c_{1}$ and $c_{2}$ are constants of integration. Equations (85) and (86) are the solutions for constant deceleration parameters. We are not interested in these two solutions but we are interested only on the first solution, which is new. For convenience, in the following we consider the solution for $k>0$ and $m>o$ and omit the integrating constant $c_{0}$ by setting $c_{0}=0$ in (84). By doing this, we also set the initial time of the universe to $t=0$. The reason for considering the solution only for $k>0$ and $m>o$ is not only for simplicity but also for compatibility with the observed universe. $k>0$ means we are dealing with increasing acceleration $q=-k<0$. Because $t=0$ and $k>0$ , is the only way to shift the deceleration parameter to values higher than $-1$ is to set $m>0$ . Under the above considerations, the equation (84) is further reduces to

\begin{equation}
R=R_{0}e^{\frac{2}{m}\tanh^{-1}\left(\frac{kt-m}{m}\right)} ~~; ~~ for ~~ k>0 ~ and ~ m\geq0
\end{equation}
From Eqs. (6) , (15) and (87), we got
\begin{equation}
A=R_{0}^{\frac{4n}{3n+1}}e^{\frac{8n}{m(3n+1)}\tanh^{-1}\left(\frac{kt-m}{m}\right)}
\end{equation}

\begin{equation}
B=R_{0}^{\frac{4}{3n+1}}e^{\frac{8}{m(3n+1)}\tanh^{-1}\left(\frac{kt-m}{m}\right)}
\end{equation}
With the help of Eqs. (88) and (89) , the metric (3) becomes
\begin{equation}
\begin{split}
ds^{2}=&R_{0}^{\frac{8n}{3n+1}}e^{\frac{16n}{m(3n+1)}\tanh^{-1}\left(\frac{kt-m}{m}\right)}(dx^{2}+dy^{2}+dz^{2})\\
&+R_{0}^{\frac{8}{3n+1}}e^{\frac{16}{m(3n+1)}\tanh^{-1}\left(\frac{kt-m}{m}\right)}d\psi^{2}-dt^{2}
\end{split}
\end{equation}
Equation (90) represents a Bianchi type-I cosmological model universe in Lyra geometry with linearly varying deceleration parameter.\\
\\
\textbf{Some Physical Properties of the Model (90) with linearly varying Deceleration Parameter}\\
\\
Equations (6)-(10), respectively, give us the Spatial Volume $V$ , Hubble's Parameter $H$ , Expansion scalar $\theta$ , Shear scalar $\sigma$ , mean anisotropy parameter $\Delta$ as follows:
\begin{equation}
V=R_{0}^{4}e^{\frac{8}{m}\tanh^{-1}\left(\frac{kt-m}{m}\right)}
\end{equation}
\begin{equation}
H=\frac{\dot{R}}{R}=\frac{2}{2mt-kt^{2}}
\end{equation}
\begin{equation}
\theta=\frac{8}{2mt-kt^{2}}
\end{equation}
\begin{equation}
\sigma^{2}=\frac{24(n-1)^{2}}{(3n+1)^{2}(2mt-kt^{2})^{2}}
\end{equation}
\begin{equation}
\Delta=\frac{3(n-1)^{2}}{(3n+1)^{2}}= ~ constant ~ (~\neq0 ~~ for ~~ n\neq1)
\end{equation}
From Eqs. (93) and (94) we have
\[\frac{\sigma^{2}}{\theta^{2}}=\frac{3(n-1)^{2}}{8(3n+1)^{2}}\]
Therefore
\begin{equation}
\lim_{t\to \infty}\frac{\sigma^{2}}{\theta^{2}}=\frac{3(n-1)^{2}}{8(3n+1)^{2}} ~\neq0 ~~ for ~~ n\neq1
\end{equation}\\
Again from (22), (11) and (13), the displacement vector $\beta$, the energy density $\rho$ and the fluid pressure $p$ are obtained as follows:
\begin{equation}
\beta=\beta_{0}e^{-\frac{8}{m}\tanh^{-1}\left(\frac{kt-m}{m}\right)}
\end{equation}
\begin{equation}
\rho=-\frac{192n(n+1)}{(3n+1)^{2}(2mt-kt^{2})^{2}}+\frac{3}{4}\beta_{0}^{2}e^{-\frac{16}{m}\tanh^{-1}\left(\frac{kt-m}{m}\right)}
\end{equation}
\begin{equation}
p=-\frac{48n(kt-m)}{(3n+1)^{2}(2mt-kt^{2})^{2}}-\frac{3}{4}\beta_{0}^{2}e^{-\frac{16}{m}\tanh^{-1}\left(\frac{kt-m}{m}\right)}
\end{equation}
\\
Variations of some parameters with respect to time for this case III are shown in Figs. 15-17 \\

\includegraphics{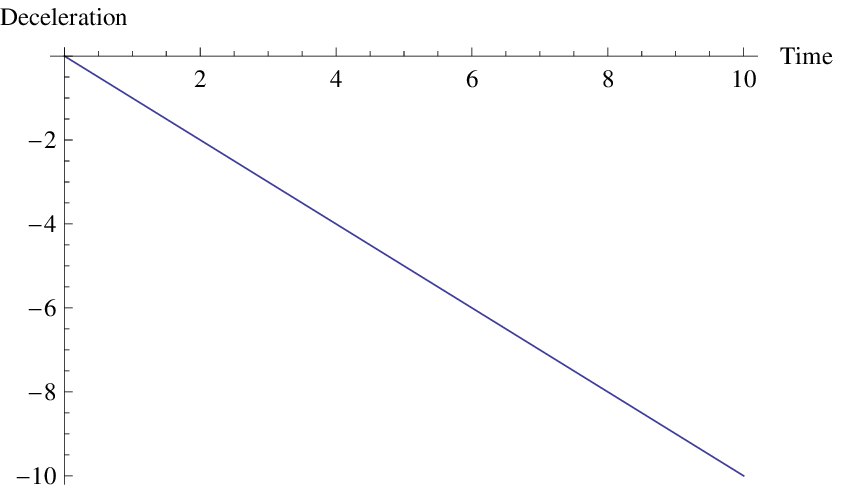}
\\
\textbf{Fig. 15 : The plot of Deceleration Parameter q vs. Time t. Here $k=1 (k > 0), m = 1 (m\geq 0)$}\\
\\
\includegraphics{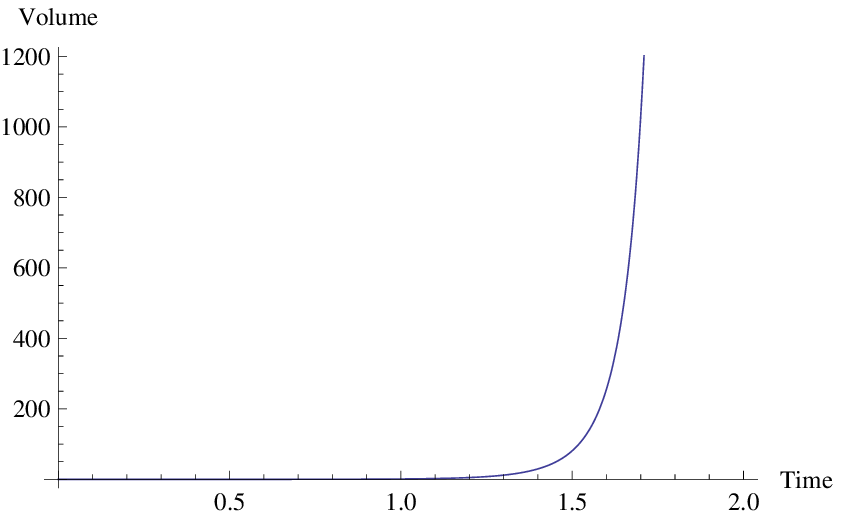}
\\
\textbf{Fig. 16 : The plot of Volume V vs. Time t. Here $k=1 (k > 0), m = 1 (m\geq 0), n=2$ and $R_0=1$}\\
\\
\includegraphics{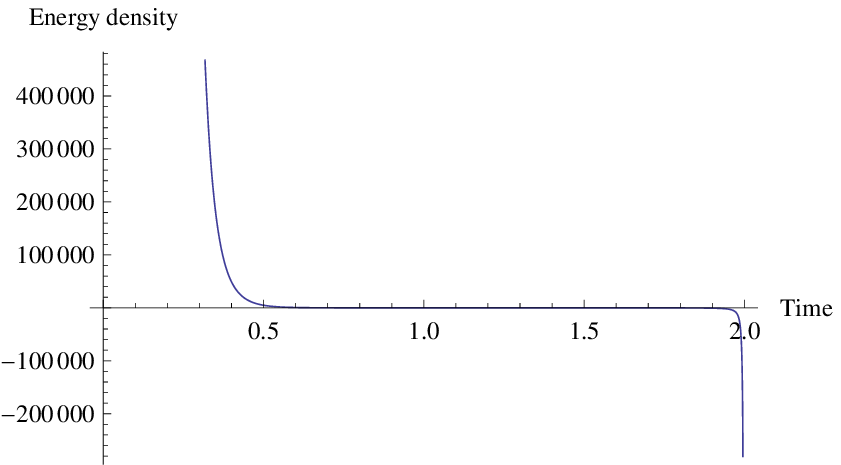}
\\
\textbf{Fig. 17 : The plot of Volume Energy density $\rho$ vs. Time t. Here $k=1 (k > 0), m = 1 (m\geq 0), n=2$ and $R_0=1$}\\
\\
\\
\textbf{Case - IV : Models with special form of Deceleration Parameter}\\
\\
In this case we have considered a special form of deceleration parameter (see [80]) as follows:
\begin{equation}
q=-\frac{R\ddot{R}}{\dot{R}^{2}}=-1+\frac{c}{1+R^{c}}
\end{equation}
where the average scale factor $R$ is a function of time $t$ and $c>0$ is a constant.\\
\\
Equation (100) gives us
\begin{equation}
R=\left(d_{2}e^{d_{1}ct}-1\right)^{\frac{1}{c}}
\end{equation}
Therefore from Eqs. (6) , (15) and (101), we have
\begin{equation}
A=\left(d_{2}e^{d_{1}ct}-1\right)^{\frac{4n}{c(3n+1)}}
\end{equation}

\begin{equation}
B=\left(d_{2}e^{d_{1}ct}-1\right)^{\frac{4}{c(3n+1)}}
\end{equation}
\\
Therefore by using Eqs. (102) and (103) , the metric (3) can be obtained as follows:
\begin{equation}
ds^{2}=\left(d_{2}e^{d_{1}ct}-1\right)^{\frac{8n}{c(3n+1)}}(dx^{2}+dy^{2}+dz^{2})+\left(d_{2}e^{d_{1}ct}-1\right)^{\frac{8}{c(3n+1)}}d\psi^{2}-dt^{2}
\end{equation}
\\
Equation (104) represents a Bianchi type-I cosmological model universe in Lyra geometry with special form of deceleration parameter.\\
\\
\textbf{Some Physical Properties of the Model (104) with special form of Deceleration Parameter}\\
\\
Equations (6)-(10), respectively, give us the Spatial Volume $V$ , Hubble's Parameter $H$ , Expansion scalar $\theta$ , Shear scalar $\sigma$ , mean anisotropy parameter $\Delta$ as follows:
\begin{equation}
V=\left(d_{2}e^{d_{1}ct}-1\right)^{\frac{4}{c}}
\end{equation}
\begin{equation}
H=\frac{\dot{R}}{R}=d_{1}d_{2}\left(d_{2}e^{d_{1}ct}-1\right)^{-1}
\end{equation}
\begin{equation}
\theta=4d_{1}d_{2}\left(d_{2}e^{d_{1}ct}-1\right)^{-1}
\end{equation}
\begin{equation}
\sigma^{2}=\frac{6(n-1)^{2}d_{1}^{2}d_{2}^{2}}{(3n+1)^{2}}\left(d_{2}e^{d_{1}ct}-1\right)^{-2}
\end{equation}
\begin{equation}
\Delta=\frac{3(n-1)^{2}}{(3n+1)^{2}}= ~ constant ~ (~\neq0 ~~ for ~~ n\neq1)
\end{equation}
From Eqs. (107) and (108), we have
\[\frac{\sigma^{2}}{\theta^{2}}=\frac{3(n-1)^{2}}{8(3n+1)^{2}}\]
Therefore
\begin{equation}
\lim_{t\to \infty}\frac{\sigma^{2}}{\theta^{2}}=\frac{3(n-1)^{2}}{8(3n+1)^{2}} ~\neq0 ~~ for ~~ n\neq1
\end{equation}\\
Again from (22), (11) and (13), the displacement vector $\beta$, the energy density $\rho$ and the fluid pressure $p$ are obtained as follows:
\begin{equation}
\beta=\beta_{1}\left(d_{2}e^{d_{1}ct}-1\right)^{-\frac{4}{c}}
\end{equation}
\begin{equation}
\rho=-\frac{48n(n+1)d_{1}^{2}d_{2}^{2}}{(3n+1)^{2}}\left(d_{2}-e^{-d_{1}ct}\right)^{-2}+\frac{3}{4}\beta_{1}^{2}\left(d_{2}e^{d_{1}ct-1}\right)^{-\frac{8}{c}}
\end{equation}
\begin{equation}
\begin{split}
p=-\frac{96n^{2}d_{1}^{2}d_{2}^{2}}{(3n+1)^{2}}\left(d_{2}-e^{-d_{1}ct}\right)^{-2}&+\frac{12nd_{1}^{2}d_{2}ce^{-d_{1}ct}}{3n+1}\left(d_{2}-e^{-d_{1}ct}\right)^{-2}\\
   &-\frac{3}{4}\beta_{1}^{2}\left(d_{2}e^{d_{1}ct-1}\right)^{-\frac{8}{c}}
\end{split}
\end{equation}
\\
Also from Eqs. (100) and (101), the deceleration parameter $q$ is found as follows:
\begin{equation}
q = -1 + \frac{1}{d_{2}}e^{-d_{1}ct}
\end{equation}
\\
Variations of the some important parameters with respect to time for case IV are shown in Figs. 18-20. \\

\includegraphics{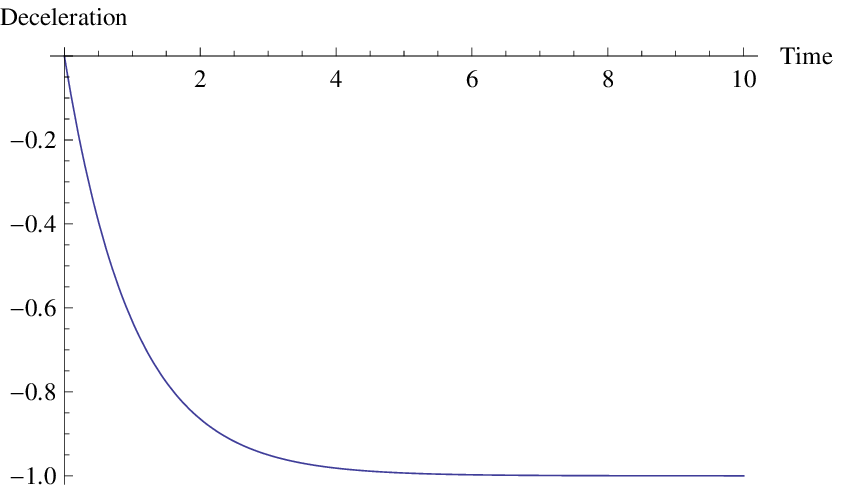}
\\
\textbf{Fig. 18 : The plot of Deceleration Parameter vs. Time t. Here $c = d_1=d_2=\beta_1=1$ and n=2}\\

\includegraphics{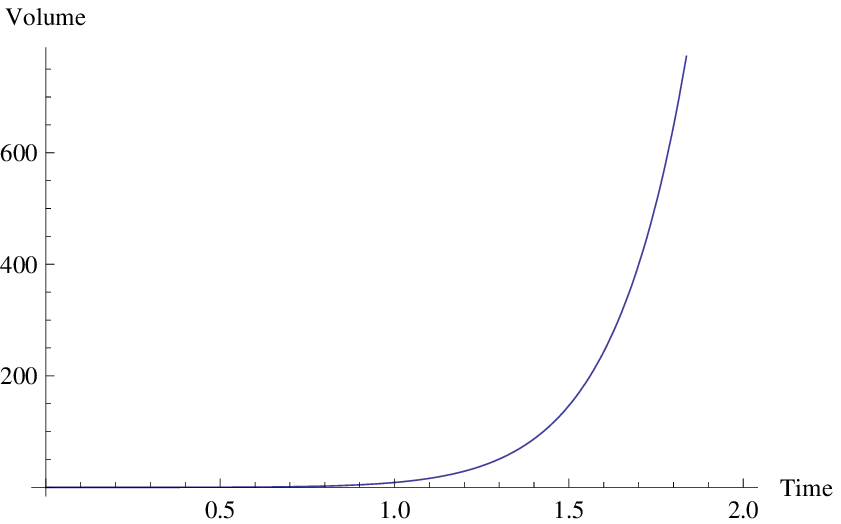}
\\
\textbf{Fig. 19 : The plot of Volume V vs. Time t. Here $c = d_1=d_2=\beta_1=1$ and n=2}\\

\includegraphics{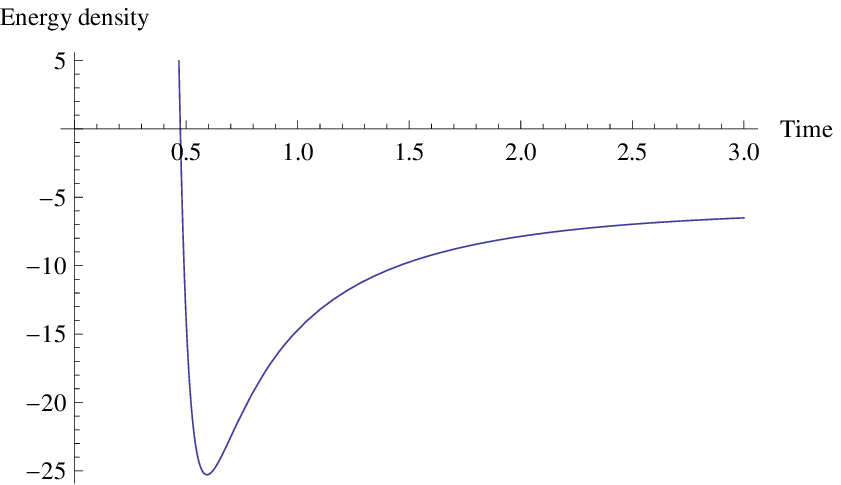}
\\
\textbf{Fig. 20 : The plot of Energy density $\rho$ vs. Time t. Here $c = d_1=d_2=\beta_1=1$ and n=2}\\

\section{Physical Interpretations of the solutions}
\label{sect:Interpretation}

\textbf{Case - I : Models with Time Dependent Deceleration Parameter}\\
\\
(i)     Equation (23) gives the displacement vector $\beta$ as $\beta=\left(k_{0}t^{a}e^{t}\right)^{-\frac{4}{b}}$ , which is a decreasing function of
        cosmic time $t$ and finally it becomes zero when $t\rightarrow\infty$.\\
        \\
(ii)    From Eqs. (24) and (25) we have seen that the energy density $\rho$ and pressure $p$ diverge at $t=0$ but when $t\rightarrow\infty$ then the energy
        density $\rho$ becomes zero whereas the pressure takes the finite value $\frac{12an}{b(3n+1)^{2}}$. \\
\\
(iii)   Again from Eq. (26) it is clear that initially when $t=0$ the spatial volume $V=0$ gradually increases with the passage of time $t$ and it becomes
        infinite when $t\rightarrow\infty$.\\
        Also from Eqs. (27) and (28) we have seen that both Hubble's parameter $H$ and expansion scalar $\theta$ are infinite at $t=0$ and expands with the increase of cosmic time. Thus our model universe is evolves at time $t=0$ and is expanding that explains the Big-Bang scenario of our model universe.\\
\\
(iv)    The average scale factor given by Eq. (16) is zero at initial epoch of time (t=0), so our model has a point type singularity.\\
\\
(v)     Also from Eqs. (17) and (27) we observed that at $t\rightarrow\infty$ the deceleration parameter $q=-1$ and $\frac{dH}{dt}=0$. Thus for
        $t\rightarrow\infty$ the Hubble's parameter $H$ is maximum and the model will be increasing at fastest rate for $t\rightarrow\infty$.\\
\\
(vi)    Equations (30) and(31) give us that the mean anisotropy parameter $\Delta$ and $\lim_{t\to \infty}\frac{\sigma^{2}}{\theta^{2}}$ are non zero
        constants for all values of time $t$ whenever $n\neq1$, so from the mathematical point of view our model will be an anisotropic one. From this result, we seem that during the process of evolution of our model universe there may have some possibilities to have an anisotropic universe for some duration. But it does not contradict to the present observational fact that our present universe is an isotropic one, because the initial anisotropy of the Bianchi Type-I universe quickly dies away and eventually it  evolves into a FRW universe as suggested by [81]. Also, a Bianchi Type-I universe has a different scale factor for each of the three spatial directions, and so anisotropy is introduced to the system automatically representing an anisotropic universe. Departures from isotropy that preserve homogeneity are described by Bianchi model with the overall geometry of space. One may show [82] that only certain Bianchi types specially type I, V, IX etc. allow for isotropic limit. But by conducting general test of isotropy using CMB temperature and polarization data from Plank, Saadeeh et al.[83] strongly disfavoured the anisotropic expansion of the universe with odds of 121000:1 against . But still some share of anisotropic is there, so we required for further investigation in this area mathematically as well as observationally.\\
        \\
        But interestingly when $n = 1$ our model universe approach to an isotropic one which is a good indication favouring the recent observational findings.\\
\\
(vii)   Again from Eq. (17) we found that as $t\rightarrow\infty$ the value of the deceleration parameter $q$ becomes $-1$. Also when we plot the
        deceleration parameter versus time as shown in \textbf{Fig. 1}, it seems that our model universe is decelerating at the initial phase and changing from decelerating to accelerating as time progresses. It means that our model universe might have undergone a transition from early deceleration to late time acceleration at certain point of time supporting the recent observational findings. Such type of situation can be stimulated due to the time varying deceleration parameter which may be positive at the early stage and evolves negative values at the late time. Here, in our problem the deceleration parameter is generated from the scale factor $R=\left(t^{a}e^{t}\right)^{\frac{1}{b}}$ which have a hybrid form containing the factors of exponential behavior and power law behavior which are widely used in the very common literatures for the investigation of background cosmology. Since the power law factor of this scale factor dominates the early part of cosmic dynamics and the exponential factor part dominates at late time providing a realistic model universe.\\
\\
(viii)  From \textbf{Fig. 2} of Eq. (26) we have seen that the that the volume $V$ is an increasing function of time $t$ and it becomes infinite as
        $t\rightarrow\infty$. Initially, the volume $V$ increases very rapidly but as the time progresses the rate of increase becomes slow.\\
\\
(ix)    The \textbf{Fig. 3} explains that the energy density of our model universe has a finite value at initial epoch of time (t=0), that is at the beginning
        of evolution; and even when $t\rightarrow\infty$ it has some finite value and there is no singularity; there by, there is possibility that our model is that of oscillating type.\\
\\
\textbf{Case - II : Models with variable Deceleration Parameter}\\
\\
\textbf{Sub Case-1 of Case-II :}\\
\\
(i)     From Eq. (38), it is clear that the average scale factor R(t) can never be negative if $k_{2}>0$. From Fig. 4, we have seen that in the early
        stages of the universe, i.e. when $t \rightarrow 0$ then the scale factor of the universe is almost constant and increasing very slowly as the time progresses. But, after a certain period of time our universe exploded suddenly and starts expansion at large scale, which is consistent with Big Bang scenario.\\
\\
(ii)    Figure 5 explains that initially the volume $V$ of the universe is constant and it increases with the passage of time and finally the
        volume becomes infinite as  $t \rightarrow \infty$.\\
\\
(iii)   From Eq. (48) and its graph i.e. Fig. 6, it has been observed that $\beta$ is always positive and is a decreasing function of time. In
        our model, displacement vector $\beta$ plays the role of cosmological constant and preserves the same character as $\Lambda$-term in the Einstein's theory of relativity, in fact with respect to the recent observations.\\
\\
(iv)    In the early stage, i.e. When $t \rightarrow 0$ then for very small values of $k_{1}$, we have seen from Eqs. (49) and (50) and their graphs, i.e.
        Figs. 7 and 8, that the energy density is a decreasing function of time and tends to a small positive quantity whereas the pressure $p$ is an increasing function of time and tends to a small negative quantity.\\
\\
(v)     Again, the Eq. (51) gives us q = -1, so that our cosmological model represents an accelerating universe.\\
\\
\textbf{Sub Case-2 of Case-II :}\\
\\
(i)     Equations (53) and (57) show that both scale factor R(t) and volume V are positive for $0 \leq t < \infty$ if $\alpha_{1}$, $\alpha_{2}$ and
        $\alpha_{3}$ are positive constants. Also Figs. 9 and 10 explain that both scale factor R(t) and volume $V$ are increasing functions of time, implying that our universe is expanding.\\
        \\
(ii)    From Fig. 11 and Eq. (63), it has been observed that $\beta$ is positive and is a decreasing function of time and  it
        tends to zero in infinite time. Characteristically $\beta$ is similar to that of $\Lambda$ term in Einstein's theory of gravity. In this model, $\beta$ also plays the same role as cosmological constant and preserves the same character as $\Lambda$ term in Einstein's theory.\\
        \\
(iii)   Again, from Eqs. (64) and (65), we have seen that both energy density and pressure are negative, but whenever $t\rightarrow\infty$ both of them
        converges to zero.\\
        \\
(iv)    From Eqs. (61) and (62), it is observed that both mean anisotropic parameter $\Delta$ and $\lim_{t\to \infty}\frac{\sigma^{2}}{\theta^{2}}$ are
        constant and non zero for all values of time $t$ whenever $n\neq1$ and $n\neq\frac{1}{3}$, so our model remains anisotropic throughout the evolution of the universe. But whenever $n = 1$ then both $\Delta$ and $\lim_{t\to \infty}\frac{\sigma^{2}}{\theta^{2}}$ become zero for all values of time $t$ so our model universe will be an isotropic universe for this particular case.\\
\\
\textbf{Sub Case-3 of Case-II :}\\
\\
(i)     From Eqs. (72) and (74), it is clear that initially when $t=0$ the spatial volume $V=0$ and gradually increases with the passage of time $t$ and it
        becomes infinite when $t\rightarrow\infty$ whereas the expansion scalar $\theta$ is infinite at $t=0$. \\
        Again at $t\rightarrow\infty$, Eqs. (73) and (81) give us $\frac{dH}{dt}=0$ and deceleration parameter $q=-1$, so the Hubble's parameter $H$ is maximum. Thus at $t\rightarrow\infty$  our model is expanding at fastest rate. Hence our model universe is evolves at time $t=0$ and is expanding, which explains the Big-Bang scenario of our model universe. \\
\\
(ii)    Equation (78) shows that the displacement vector $\beta=[\sinh(k_{5}t)]^{-\frac{4}{n}}$ is a decreasing function of cosmic time $t$ and finally it
        becomes zero when $t\rightarrow\infty$.\\
\\
(iii)   From Eqs. (79) and (80) we have seen that the energy density $\rho$ and pressure $p$ diverge at $t=0$ and tend to zero at $t\rightarrow\infty$.\\
\\
(iv)    The average scale factor given by Eq. (68) and the spatial scale factors given by Eqs. (69) and (70) are all zero at initial epoch of time
        $t=0$, so our model has a point type singularity.\\
\\
(v)     Equations (76) and (77) give us that the mean anisotropy parameter $\Delta$ and $\lim_{t\to \infty}\frac{\sigma^{2}}{\theta^{2}}$ are non zero
        constants for all values of time $t$ whenever $n\neq1$, so our model remains anisotropic throughout the evolution of the universe and our model universe does not approach to isotropy.\\
\\
(vi)    Again from Eq. (81) and from the graph of deceleration parameter vs. time given by Fig. 12, it is clear that the deceleration parameter
        $q$ in this case also changes sign from positive to negative, so we can conclude that initially our model universe is decelerating at initial phase and then changes from decelerating to accelerating. \\
\\
(vii)   Also from Fig. 14 of equation (79), it is seen that initially our model universe has infinite density and as the time progresses it becomes
        negative and finally energy density tend to a negative finite value. Thus this model universe approaches to a steady state.\\
\\
\\
\textbf{Case - III : Models with Linearly Varying Deceleration Parameter}\\
\\
(i)     From Eqs. (91) and (33), it is clear that initially when $t=0$ the spatial volume $V$ has a constant value $V=R_{0}^{4}$ and gradually increases
        with the passage of time $t$ and it becomes infinite when $t\rightarrow\infty$ whereas the expansion scalar $\theta$ is a decreasing function of time $t$ which is infinite at $t=0$ and finally it becomes zero when $t\rightarrow\infty$. \\
        Equations (92)-(94) imply that the Hubble's parameter $H$, expansion scalar $\theta$ and shear scalar $\sigma$ diverge for $t=\frac{2m}{k}$.\\
        \\
(ii)    The model (90) does not have any singularity, because the spatial scale factors $A$ and $B$ given by Eqs. (88) and (89), respectively, are non zero
        for all values of $t$.\\
        \\
(iii)   From Eq. (82), it is seen that initially when $t=0$ then $q=m-1$ so that the model universe is decelerating for $m>1$ and accelerating for
        $0 \leq m \leq 1$. But for $t>\frac{m-1}{k}$ , $q>0$, the model universe starts accelerating expansion and whenever $t=\frac{m}{k}$ , $q=-1$ implying that the universe experiences super exponential expansion and ends with $q=-m-1$ at $t=\frac{2m}{k}$.\\
        \\
(iv)    From Eqs. (95) and (96) we have the mean anisotropy parameter $\Delta=\frac{3(n-1)^{2}}{(3n+1)^{2}}= ~ constant ~ (~\neq0 ~~ for ~~ n\neq1)$ and
        $\lim_{t\to \infty}\frac{\sigma^{2}}{\theta^{2}}=\frac{3(n-1)^{2}}{8(3n+1)^{2}} ~(\neq0 ~~ for ~~ n\neq1)$ so the model is anisotropic throughout the evolution of the universe except at $m=1$ , hence the model does not approach to isotropy.\\
        \\
 (v)    From Eq. (97) we have also got that the displacement vector $\beta$ is a decreasing function of time $t$ and it has a finite value at the initial
        epoch of time i.e. at $t=0$. \\
        \\
(vi)    Equations (98) and (99) give that both the energy density  $\rho = -\infty$ and pressure $p=-\infty$ for $t=0$ and they will take finite value
        $\frac{3}{4}\beta_{0}^{2}$ as $t\rightarrow\infty$. \\
        \\
(vii)   Again from Eq. (82) and Fig. 15, it is clear that the deceleration parameter $q$ changes sign from positive to negative, so we can
        conclude that initially our model universe is decelerating (at initial phase) and then changes from decelerating to accelerating. \\
\\
Hence this model universe is also consistent with the result of recent cosmological observations.\\
\\
\textbf{Case - IV : Models with special form of Deceleration Parameter}\\
\\
(i)     From Eqs. (105) and (107), it is clear that initially when $t=0$ the spatial volume $V$ has a finite value $V=(d_{2}-1)^{\frac{4}{c}}$ and gradually
        increases with the passage of time $t$ and it becomes infinite when $t\rightarrow\infty$ whereas the expansion scalar $\theta$ is a decreasing function of time $t$ which is finite at $t=0$ and finally it becomes zero when $t\rightarrow\infty$. \\
        \\
(ii)    The model (101) does not have any singularity, because the spatial scale factors $A$ and $B$ given by Eqs. (102) and (103) respectively are non zero
        for all values of $t$.\\
        \\
(iii)    From Eqs. (109) and (110), we have the mean anisotropy parameter $\Delta=\frac{3(n-1)^{2}}{(3n+1)^{2}}= ~ constant ~ (~\neq0 ~~ for ~~ n\neq1)$ and
        $\lim_{t\to \infty}\frac{\sigma^{2}}{\theta^{2}}=\frac{3(n-1)^{2}}{8(3n+1)^{2}} ~(\neq0 ~~ for ~~ n\neq1)$ so the model is anisotropic throughout the evolution of the universe except at $n=1$ , hence the model does not approach to isotropy.\\
        \\
(iv)     From Eq. (111) we have got that the displacement vector $\beta$ is a finite quantity at $t=0$ and become zero as $t\rightarrow\infty$. \\
        \\
(v)     Equations (112) and (113) give that both the energy density  $\rho = -\infty$ and pressure $p=-\infty$ for $t=0$ and they will take finite values as
        $t\rightarrow\infty$. \\
        \\
(vi)    Lastly in this case also from Eq (114) and the graph shown in  Fig. 18 it is clear that our model universe is an accelerating one. \\
\\
       From all the above different cases we find that the energy density and gauge function are positive and decreasing function of time. We also observed the value of deceleration parameter obtained from all the cases of our model universe are in fair agreement with the result of cosmological observations like Type SNeIa supernova, CMB anisotropies, the large scale galaxies structures of universe, Baryon Acoustic Oscillations, WMAP, and new data sets like Planck results, ACT and SPT which have measured the CMB temperature and polarisation anisotropies. From the values of pressure $p$ and critical density $\rho$ we found that our model universe behaves as the dark energy model universe which is accelerated at the late phase of the cosmic dynamics before it might be decelerating at the early phase.\\
\\
\section{Conclusion}	

It may be a good idea to try to search for the hidden source of the dark energy which dominates the universe with positive energy density and negative pressure, and responsible to produce sufficient acceleration in late time evolution of the Universe. So within the frame work of Lyra Geometry while investigating five dimensional LRS Bianchi type-I model universe with time dependent deceleration parameter interacting with vector field $\phi_{i}$, interestingly we found the above model universes behaves as a dark energy model universes which are consistent with the observational findings. Here the displacement field, which is consider as a component of total energy, plays the role of dark energy. In all the cases which we consider here with are expanding and anisotropic through the evolution supporting the present day observational findings. Thus it is seen that our models behaves as a dark energy field universe. From our findings, we seem that Lyra manifold itself contribute to dark energy consistent with the recent cosmological observations. Further study of such type of universe will be helpful for explaining the present accelerated expansion behavior of the universe. \\
\\


\begin{thebibliography}{0} 

\bibitem{jpap} A. G. Riess et al., Observational Evidence from Super-Novae for an Accelerating Universe and a Cosmological Constant, {\it Astronom. J.}, {\bf
        116}, (1988) 1009;\\
        A. G. Riess et al., Type Ia Supernova Discoveries at $z > 1$ From the Hubble Space Telescope: Evidence for Past Deceleration and Constraints on Dark Energy Evolution1, {\it Astrophys. J.} {\bf 607}, (2004) 665.\\
        S.Perlmutter,  et al., Measurements of Omega and Lambda from 42 High-Redshift Supernova {\it Astrophys. J.} {\bf 517}, (1999) 565.
\bibitem{jpap} R. Amanullah, et al., Spectra and Hubble Space Telescope Light Curves of Six Type Ia Supernovae at $0.511 < z < 1.12$ and the Union2 Compilation,
        {\it Astrophys. J.}, {\bf 716}, (2010) 712.
\bibitem{colla} P. Astier, et al., The Supernova Legacy Survey: measurement of $\Omega_{m}$, $\Omega_{\Lambda}$ and $w$ from the first year data set {\it Astron.
        Astrophys.} {\bf 447}, (2006) 31;\\
        N. Suzuki et al., "The Hubble Space Telescope Cluster Supernova Survey : V. Improving the Dark Energy Constraints Above $Z > 1$ and Building an
        Early-Type-Hosted Supernova Sample {\it Astrophys. J.} {\bf 746}, (2012) 85, (27 pp); arXiv:1105.3470v1 [astro-ph.CO] 17 May 2011.
\bibitem{jpap} C. L. Bennett, et al., First Year Wilkinson Microwave Anisotropy Probe (WMAP1 ) Observations: Preliminary Maps and Basic Results {\it Astrophys.
        J.Suppl.} {\bf 148} (2003) 1.
\bibitem{jpap} D. N. Spergel, et al., First-Year Wilkinson Microwave Anisotropy  Probe $(WMAP)^{1}$ Observations : Determination of Cosmological Parameters {\it
        Astrophys. J. Suppl.} {\bf 148} (2003) 175.
\bibitem{jpap} S.F. Daniel, et al., Large Scale Structure as a Probe of Gravitational Slip {\it Phys. Rev. D.} {\bf 77} (2008) 103513;
        http://dx.doi.org/10.1103/PhysRevD.77.103513 ;\\
        S.W. Allen  et al., Constraints on dark energy from Chandra observations of the largest relaxed galaxy clusters {\it Mon. Not. R. Astron. Soc.}
        {\bf 353} (2004) 457.
\bibitem{jpap} L. Anderson et al., The clustering of galaxies in the SDSS-III Baryon Oscillation Spectroscopic Survey: Baryon Acoustic Oscillations in the Data
        Release 9 Spectroscopic Galaxy Sample {\it Mon. Not. R. Astron. Soc.} {\bf 427} (2013) 3435; arXiv:1203.6594v1 [astro-ph.CO] 29 Mar 2012 ;\\
        D. J. Eisenstein et al., Detection of the baryon acoustic peak in the large-scale correlation function of SDSS luminous red galaxies, {\it Astrophys. J.} {\bf 633} (2005)  560.
\bibitem{jpap} E. Komatsu et al., Five-Year Wilkinson Mickrowave Anisotropy Probe ($WMAP^{1}$) Observations : Cosmological Interpretation {\it Astrophys. J.
        Suppl.} {\bf 180} (2009) 330; \\
        G. Hinshaw et al., Nine-Year Wilkinson Microwave Anisotropy Prob (WMAP) Observations : Cosmological Parameter Results {\it Astrophys. J. Suppl.} {\bf 208} (2013) 19 (25 pp), doi: 10.1088/0067-0049/208/2/19.
\bibitem{jpap} A. J. Nishizawa, Integrated Sachs Wolfe Effect and Rees Sciama Effect  {\it Prog. Theor. Exp. Phys.} {\bf 2014} (6) (2014) 06B110;
        arXiv:1404.5102v1 [astro-ph.CO] 21 Apr 2014
\bibitem{jpap} U.Seljak  et al., Cosmological parameter analysis including SDSS Ly $\alpha$ forest and galaxy bias: Constraints on the primordial spectrum of
        fluctuations, neutrino mass, and dark energy" {\it Phys. Rev. D} {\bf 71} (2005) 103515;\\
        M. Tegmark et al., Cosmological parameters from SDSS and WMAP {\it Phys. Rev. D} {\bf 69} (2004)  103501; arXiv:astro-ph/0310723v2 15 Jan 2004 ;\\
        K. Abazajian et al., The Second Data Release of the Sloan Digital Sky Survey {\it Astron. J.} {\bf 128} (2004)  502; arXiv:astro-ph/0403325v1 ;\\
        J.K. Adelman McCarthy  et al., The Sixth Data Release of the Sloan Digital Sky Survey {\it Astrophys. J. Suppl. Ser.} {\bf 175} (2) (2008)  297.
\bibitem{jpap} Plank Collab. (P. A. R. Ade et al.), Planck 2015 results. XIII. Cosmological parameters, {\it Astron. Astrophys.} {\bf June 20},  (2016);
        arXiv:1502.01589v3 [astro-ph.CO] 17 Jun 2016 ;\\
        P. A. R. Ade et al., Planck 2013 results. XVI. Cosmological parameters, {\it Astron. Astrophys.} {\bf 571} (2014) A16.
\bibitem{jpap} J. L. Sievers et al., THE ATACAMA COSMOLOGY TELESCOPE: COSMOLOGICAL PARAMETERS FROM THREE SEASONS OF DATA; preprint (2013), arXiv:1301.0824v3
        [astro-ph.CO] 11 Oct 2013; \\
\bibitem{jpap}	K. T. Story et al., A MEASUREMENT OF THE COSMIC MICROWAVE BACKGROUND DAMPING TAIL FROM THE 2500-SQUARE-DEGREE SPT-SZ, preprint (2013);
        arXiv:1210.7231v2 [astro-ph.CO] 9 Dec 2013\\
\bibitem{jpap} T. Padmanabhan and T. R. Choudhury, Can the clustered dark matter and the smooth dark energy arise from the same scalar field?, {\it Phys. Rev.D.}
        {\bf 66} (2002) 081301.\\
\bibitem{jpap} M. Elmardi, A. Abebe and A. Tekola, Chaplygin-gas solutions of f(R) gravity, {\it Int. J. Geom. Methods Mod. Phys.} {\bf 13} (2016) 1650120 (11
        pp).\\
\bibitem{jpap} M. P. Dabrowski , Phantom dark energy and its cosmological consequences,  preprint (2007), arXiv:gr-qc/0701057v1 10 Jan 2007
\bibitem{jpap} R. de Putter and E. V. Linder, Kinetic k-essence and Quintessence, {\it Astropart. Phys.}, {\bf 28} (2007) 263-272.
\bibitem{autbk} H. Weyl, {\it Vorlesungen  \"{u}ker Allgemeine relativitotstheorie}, Sitzungsberichte Der Preussischen Academie Der Wissenschaften,
        Sitzungsberichte, Zu Berlin, (1918) 465pp.
\bibitem{jpap} G. Lyra,  \"{U}ber-eine Modifikation der Riemannschen Geometrie {\it Math. Z.} {\bf 54} (1) (1951) 52, doi:10.1007/BF01175135.
\bibitem{jpap} C. Brans and R.H. Dicke ,  Mach's Principle and a Relativistic Theory of Gravitation {\it Phys. Rev.} {\bf 124} (1961) 925.
\bibitem{jpap} B.M. Barker ,  General scalar-tensor theory of gravity with constant G {\it Astrophys. J.} {\bf 219} (1978) 5 .
\bibitem{jpap} J.D. Bekenstein,   Relativistic gravitation theory for the modified Newtonian dynamics paradigm  {\it Phys. Rev. D} {\bf 70} (2004) 083509 ,
        arXiv:astro-ph/0403694v6 .
\bibitem{jpap} H. Weyl , Gravitation and electricity {\it Sitzungsber. Preuss. Akad. Wiss. Berlin (Math. Phys.)} {\bf 1918} (1918) 465-480.
\bibitem{jpap}	S. Nojiri et al., Modified gravity with negative and positive powers of the curvature: unification of the inflation and of the cosmic
        acceleration, preprint (2003), arXiv:hep-th/0307288v4 19 Sep 2003 ;\\
        T. P. Sotiriou, f(R) theories of gravity, preprint (2010), arXiv:0805.1726v4 [gr-qc] 4 Jun 2010 ;\\
        S. Nojiri et al., Unified cosmic history in modified gravity: from F(R) theory to Lorentz non-invariant models, preprint (2011), arXiv:1011.0544v4 [gr-qc] 29 May 2011 \\
        H.A. Buchdahl,  Non-Linear Lagrangians and Cosmological Theory, {\it Mon. Not. R. Astron. Soc.}, {\bf 150} (1970) 1-8.
\bibitem{jpap} A. H. Chamseddine et al., Mimetic Dark Matter, preprint (2013), arXiv:1308.5410v1 [astro-ph.CO] 25 Aug 2013
\bibitem{jpap} S. Nojiri et al., Mimetic F(R) gravity: inflation, dark energy and bounce, preprint (2014), arXiv:1408.3561v3 [hep-th] 18 Dec 2014 ;\\
        R. Myrzakulov et al., Inflation in f(R, $\phi$)-theories and mimetic gravity scenario, preprint (2015), arXiv:1504.07984v3 [gr-qc] 17 Nov 2015
\bibitem{jpap} E. Scheibe, \"{U}ber-einen verallgemeinerten affinen Zusammenhang, {\it Math. Z.}, {\bf 57} (1952) 65-74; doi:10.1007/BF01192916.
\bibitem{jpap} F. Hoyle ,  A New Model for the Expanding Universe, {\ Mon. Not. R. Astron. Soc.}, {\bf 108} (1948) 372-382.
\bibitem{jpap} F. Hoyle , J.V. Narlikar ,  Mach's Principle and the Creation of Matter, {\it Proc. R. Soc. Lond. Ser. A}, {\bf 273} (1963) 1-11.
\bibitem{jpap} F. Hoyle , J.V. Narlikar , A New Theory of Gravitation, {\it Proc. R. Soc. Lond. Ser. A}, {\bf 282} (1964) 191-207.
\bibitem{jpap} H. H. Soleng ,  Cosmologies based on Lyra's geometry, {\it Gen. Relativ. Gravit.}, {\bf 19} (1987) 1213, ; doi:10.1007/BF00759100 .
\bibitem{jpap} D. K. Sen, A Static Cosmological Model, {\it Z. Phys.}, {\bf 149} (3) (1957)  311.
\bibitem{jpap} D. K. Sen and K. A. Dunn, A Scalar-Tensor Theory of Gravitation in a Modified Riemannian Manifold, {\it J. Math. Phys.}, {\bf 1216} (4)
        (1971) 578; doi:10.1063/1.1665623.
\bibitem{jpap} N. Rosen, The Bimetric Weyl-Dirac Theory and the Gravitational Constant, {\it Found. Phys.}, {\bf 13} (3) (1983)  363.
\bibitem{jpap} W. D. Halford, Cosmological theory based on Lyra's geometry, {\it Aust. J. Phys.}, {\bf 23} (1970) 863.
\bibitem{jpap}  W. D. Halford, Scalar-Tensor Theory of Gravitation in a Lyra Manifold, {\it J. Math. Phys.}, {\bf 13} (1972) 1699, doi:10.1063/1.1665894).
\bibitem{jpap} F. Rahaman, S. Chakraborty, N. Begum, M. Hussain and M. Kalam, Bianchi-IX String Cosmological Model in Lyra Geometry, {\it Pramana J. Phys.}, {\bf
        60} (6) (2003) 1153.
\bibitem{jpap} F. Rahaman, N. Begum, G. Bag and B. C. Bhui, Cosmological Model with Negative Constant Deceleration Parameter in Lyra Geometry, {\it Astrophys.
        Space Sci.}, {\bf 299} (2005) 211.
\bibitem{jpap} R. Casana, C.A.M. de Melo  and B. M. Pimental, Electromagnetic Field in Lyra Manifold: A First Order Approach, {\it Braz. J. Phys.}, {\bf 35}
        (4B) (2005)  211.
\bibitem{jpap} R. Casana, C.A.M. de Melo  and B. M. Pimental, Spinoral Field and Lyra Geometry, {\it Astrophys. Space Sci.}, {\bf 305} (2006) 125.
\bibitem{jpap} G. Mohanty, G. C.  Samanta and K. L. Mahanta, Higher Dimensional String Cosmological Model with Bulk Viscous Fluid in Lyra Manifold, {\it
        Comm. Phys.}, {\bf 17} (4) (2007) 213.
\bibitem{jpap} G. Mohanty, R.R.  Sahoo and B.K. Bishi,  Non-existence of five dimensional string cosmological models in Riemannian and Lyra geometries, {\it
        Astrophys. Space Sci.}, {\bf 319} (2009)  75.
\bibitem{jpap} K. L. Mahanta and A. K. Biswal, String Cloud and Domain Walls with Quark Matter in Lyra Geometry, {\it J. Mod. Phys.}, {\bf 3} (2012) 1479.
\bibitem{jpap} A. Asgar and M. Ansary, Bianchi Type-V Universe with Anisotropic Dark Energy in Lyra's Geometry, {\it The African Rev. Phys.}, {\bf 9}
        (2014) 0019.
\bibitem{jpap} M. R. Mollah et al., Five Dimensional String Universes in Lyra Manifold, {\it Int. J. Astron. Astrophys.}, {\bf 5} (2015) 90.
\bibitem{jpap} M. R. Mollah and K. Priyokumar,  Higher dimensional Cosmological Model Universe with Quadratic Equation of State in Lyra Geometry,
        {\it Prespacetime J.}, {\bf 7} (3) (2016)  499.
\bibitem{jpap} R. N. Henriksen and P. S. Wesson, Self-similar space-times, {\it Astrophys. Space Sci.}, {\bf 53} (1978) 499, doi: 10.1007/BF00645031.
\bibitem{jpap} G. Mohanty and G. C. Samanta, Five Dimensional Axially Symmetric String Cosmological Models with Bulk Viscous Fluid, {\it Int. J. Theor. Phys.},
        {\bf 48} (2009) 2311; doi: 10.1007/s10773-009-0020-3.
\bibitem{jpap} V. U. M. Rao and D. Neelima, Axially symmetric space-time with strange Quark matter attached to string cloud in self creation theory and general
        relativity, {\it Int. J. Theor Phys.}, {\bf 52} (2013) 354.
\bibitem{jpap} M. E. Cahill and A. H. Taub, Spherically Symmetric Similarity Solutions of the Einstein Field Equations for a Perfect Fluid, {\it Comm. Math.
        Phys.}, {\bf 21} (1) (1971) 1.
\bibitem{jpap} V. K. Shchigolev, Cosmology with an Effective $\Lambda$-Term in Lyra Manifold, preprint (2013) ; arXiv : 1307.1866v1 [gr-qc] 7 Jul (2013)
\bibitem{jpap} H. Hova, $\Lambda$CDM and Power-Law Expansion in Lyra's Geometry, preprint (2013) ; arXiv : 1303.2336v3 [gr-qc] 7 May (2013)
\bibitem{jpap} A. T. Ali and F. Rahaman, New Class of Magnetized Inhomogeneous Bianchi Type-I Cosmological Model with Variable Magnetic Permeability in Lyra
        Geometry, preprint (2013); arXiv : 1306.5739v2 [gr-qc] 11 Dec (2013)
\bibitem{jpap} M. A. Megied, R. M. Gad and E. A. Hegazy, Inhomogeneous Bianchi type-I Cosmological Model with Electromagnetic Field in Lyra Geometry, preprint
        (2014); arXiv : 1411.5978v1 [gr-qc] 21 Nov (2014)
\bibitem{jpap} M. Khurshudyan et al., Interacting Ricci Dark Energy Models with an Effective $\Lambda$-term in Lyra Manifold, preprint (2014); arXiv : 1402.5678v1 [gr-qc] 23 Feb (2014a)	
\bibitem{jpap} M. Khurshudyan, et al., Interacting Quintessence Dark Energy Models in Lyra Manifold, preprint (2014); arXiv : 1404.2141v3 [gr-qc] 18 Aug (2014b)
\bibitem{jpap} H. Saadat, A cosmological Model of the Early Universe Based on ECG with Variable $\Lambda$-term in Lyra Geometry, preprint (2015); arXiv : 1508.06544v1 [gr-qc] 24 Aug (2015)
\bibitem{jpap} F. Darabi, Y. Heydarzade and F. Hajkarim, Stability of Einstein Static Universe over Lyra Geometry, preprint (2015); arXiv : 1406.7636v2 [gr-qc] 4 Dec (2015)
\bibitem{jpap} A. H. Ziaie, A. Ranjbar and H. R. Sepangi, Trapped Surfaces and Nature of Singularities in Lyra's Geometry, preprint (2013); arXiv : 1306.2601v2 [gr-qc] 10 Jan (2015).
\bibitem{jpap} M. L. Pucheu, F. A. P. Alves Junior, A. B. Barreto  and C. Romero, Cosmological models in Weyl geometrical scalar-tensor theory, preprint (2016); arXiv : 1602.06966v1[gr-qc] 22 Feb (2016)
\bibitem{jpap} K. Priyokumar,  and M. R. Mollah, Higher Dimensional LRS Bianchi type - I Cosmological Model Universe Interacting with Perfect Fluid in Lyra
        Geometry, {\it The African Rev. Phys.}, {\bf 11} (33) (2016) 0006.
\bibitem{jpap} R. Bali and P. Kumawat, Bianchi Type I Tilted Cosmological Model for Barotropic Perfect Fluid Distribution with Heat Conduction in General
        Relativity, {\it Braz. J. Phys.}, {\bf 40} (3) (2010), doi: 10.1590/S0103-97332010000300001.
\bibitem{jpap}  A. Pradhan, B. Saha and V. Rikhvitsky, Bianchi type-I transit cosmological models with time dependent gravitational and cosmological constants
        reexamined, preprint (2013); arXiv:1308.4842v3 [physics.gen-ph] 5 Aug 2015.
\bibitem{jpap} A. J. Accioly, Exact Kantowski-Sachs and Bianchi Types I and III Cosmological Models with a Conformally Invariant Scalar Field, {\it Rev.
        Brasileira Fisica} {\bf 15} (2) (1985) 167.
\bibitem{jpap} S. P. Kandalkar, P. P. Khade and S. P. Gawande, Homogeneous Bianchi Type-I cosmological model filled with viscous fluid with a varying $\lambda$ ,
        {\it Rom. J. Phys.}, {\bf 54} (1-2) (2009) 195.
\bibitem{jpap} K. S. Adhav, R. P. Wankhade and A. S. Bansod, LRS Bianchi Type-I Cosmological Model with Anisotropic Dark Energy and Special Form of Deceleration
        Parameter, {\it J. Mod. Phys.}, {\bf 4} (2013)  1037.
\bibitem{jpap} V. U. M.Rao and D. Neelima, LRS Bianchi Type-I Dark Energy Cosmological Models in General Scalar Tensor Theory of Gravitation, {\it ISRN Astron.
        Astrophys.}, {\bf 2013} (2013), Article ID 174741, 1; doi: 10.1155/2013/174741.
\bibitem{jpap} K. S. Adhav, LRS Bianchi Type-I Universe with Anisotropic Dark Energy in Lyra Geometry, {\it Int. J. Astron. Astrophys.}, {\bf 1} (2011) 204.
\bibitem{jpap} Pradhan, A. and A. K. Singh, Anisotropic Bianchi type-I string cosmological models in normal gauge for Lyra's manifold with constant
        deceleration parameter, {\it Int. J. Theor. Phys.}, {\bf 50} (3) (2011)  916; doi: 10.1007/s10773-010-0636-3.
\bibitem{jpap} A. Asgar and M. Ansary, Exact Solutions of Axially Symmetric Bianchi Type-I Cosmological Model in Lyra Geometry, {\it IOSR J. Appl.
        Phys.}, {\bf 5} (6) (2014) (01).
\bibitem{jpap} A. Banerjee et al., Bianchi type-I cosmological models with viscous fluid in higher dimensional space time, {\it Astrophys. J.}, {\bf 358}
        (1990) 23.
\bibitem{jpap} K. D. Krori, T. Chaudhuri and C. R. Mahanta, Strings in Some Bianchi Type Cosmologies, {\it Gen. Relativ. Gravit.}, {\bf 26} (3) (1994) 265.
\bibitem{jpap} O. Gron, Inflationary Cosmology According to Wesson's Gravitational Theory {\it Astron. Astrophys.} {\bf 193} (1988) 1.
\bibitem{jpap} C. B. Collins, E. N. Glass and D. A. Wilkinson, Exact Spatially Homogeneous Cosmologies, {\it Gen. Relativ. Gravit.}, {\bf 12} (1980) 805;
        doi: 10.1007/BF00763057 .
\bibitem{jpap} B. Saha, H. Amirhashchi and A. Pradhan, Two-Fluid Scenario for Dark Energy Models in an FRW Universe-Revisited, {\it Astrophys. Space Sci.}, {\bf
        342} (2012) 257; doi: 10.1007/s10509-012-1155-x.
\bibitem{jpap} B. Mishra and S. K. Tripathy, Anisotropic dark energy model with a hybrid scale factor, preprint (2015), arXiv:1507.03515v1 [physics.gen-ph] 29 May 2015
\bibitem{jpap} O. Akarsu, S. Kumar, R. Myrzakulov, M. Sami and L. Xu, Cosmology with hybrid expansion law: scalar field reconstruction of cosmic history and
        observational constraints, {\it J. Cosmol. Astropart. Phys.}, {\bf  01} (2014) 22.
\bibitem{jpap} A. Pradhan, J. P. Shahi and C. B. Singh, Cosmological Models of Universe with Variable Deceleration Parameter in Lyra's Manifold, {\it Braz. J.
        Phys.} {\bf 36} (2006) 1227; doi: 10.1590/S0103-97332006000700020.
\bibitem{jpap} O. Akarsu and T. Dereli, Cosmological Models with Linearly Varying Deceleration Parameter {\it Int. J. Theor. Phys.} {\bf 51} (2011) 612;
        doi: 10.1007/s10773-011-0941-5.
\bibitem{jpap} A. K. Singha and U. Debnath, Acceleration Universe with a Special Form of Deceleration Parameter {\it Int. J. Theor. Phys.} {\bf 48} (2008) 351;
        doi: 10.1007/s10773-008-9807-x.
\bibitem{jpap} K. C. Jacobs, Spatially homogeneous and euclidean cosmological models with shear, {\it Astrophys. J.} {\bf 153} (2) (1968) 661 - 678.
\bibitem{jpap} G.F.R. Ellis and M.A.H. MacCallum , A class of homogeneous cosmological models {\it Comm. Math. Phys } {\bf 12} (1969)  108.
\bibitem{jpap} D.Saadesh et al., How isotropic is the Universe, preprint (2016); arXiv : 1605.0718v2 [astro-ph] 7 Sep (2016).
\end{thebibliography}
\end{document}